\documentclass[twocolumn]{emulateapj}
\usepackage{epsfig}
\usepackage{longtable}

\newcommand{\Msun}{M_{\odot}}

\shorttitle{LRG-Selected Beams}
\shortauthors{Ammons et al.}
\journalinfo{Accepted for publication in the Astrophysical Journal 11/02/13 }
\submitted{}
\begin{document}

\title{Mapping Compound Cosmic Telescopes Containing Multiple, Projected Cluster-Scale Halos}

\author{S. Mark Ammons\altaffilmark{1}, Kenneth C. Wong\altaffilmark{2, \dag}, Ann I. Zabludoff\altaffilmark{3}, Charles R. Keeton\altaffilmark{4}}

\altaffiltext{1}{Lawrence Livermore National Laboratory, Physics Division L-210, 7000 East Ave., Livermore CA 94550, ammons1@llnl.gov}
\altaffiltext{2}{Institute of Astronomy and Astrophysics, Academia Sinica (ASIAA), Taipei 10641, Taiwan, kwong@as.arizona.edu}
\altaffiltext{\dag}{EACOA Fellow}
\altaffiltext{3}{Steward Observatory, University of Arizona, 933 Cherry Ave., Tucson, AZ  85721, aiz@email.arizona.edu}
\altaffiltext{4}{Department of Physics and Astronomy, Rutgers University, 136 Frelinghuysen Road, Piscataway, NJ 08854, keeton@physics.rutgers.edu}

\begin{abstract}
Lines of sight with multiple, projected, cluster-scale gravitational lenses have high total masses and complex lens plane interactions that can boost the area of magnification, or \'etendue, making detection of faint background sources more likely than elsewhere.  To identify these new ``compound'' cosmic telescopes, we have found directions in the sky with the highest integrated mass densities, as traced by the projected concentrations of Luminous Red Galaxies (LRGs). We use new galaxy spectroscopy to derive preliminary magnification maps for two such lines of sight with total mass exceeding $\sim3 \times 10^{15} M_{\odot}$.  From 1151 MMT Hectospec spectra of galaxies down to $i_{AB} = 21.2$, we identify 2-3 group- and cluster-scale halos in each beam.   These are well-traced by LRGs.  The majority of the mass in beam J085007.6+360428 (0850) is contributed by Zwicky 1953, a massive cluster at $z=0.3774$, whereas beam J130657.5+463219 (1306) is composed of three halos with virial masses of  $6\times10^{14} $ - $2\times10^{15} M_{\odot}$, one of which is Abell 1682.  The magnification maps derived from our mass models based on spectroscopy and SDSS photometry alone display substantial \'{e}tendue:  the 68\% confidence bands on the lens plane area with magnification exceeding $10$ for a source plane of $z_s = 10$ are [1.2, 3.8] square arcminutes for 0850 and [2.3, 6.7] square arcminutes for 1306.  In deep Subaru Suprime-Cam imaging of beam 0850, we discover serendipitously a candidate multiply-imaged V-dropout source at $z_{phot}=5.03$.  The location of the candidate multiply-imaged arcs is consistent with the critical curves for a source plane of $z=5.03$ predicted by our mass model.  Incorporating the position of the candidate multiply-imaged galaxy as a constraint on the critical curve location in 0850 narrows the 68\% confidence band on lens plane area with $\mu > 10$ and $z_s = 10$ to [1.8, 4.2] square arcminutes, an \'{e}tendue range comparable to that of MACS 0717+3745 and El Gordo, two of the most powerful known single cluster lenses.  The significant lensing power of our beams makes them powerful probes of reionization and galaxy formation in the early Universe.
\end{abstract}


\section{INTRODUCTION}

Constraining the properties of high-redshift galaxies is the first step towards identifying the sources of reionizing photons and understanding the initial stages of the formation of galaxies.  Progress in this arena is limited by the faintness of high-redshift galaxies.  One productive technique for selecting high-redshift galaxies is the Lyman break dropout method in blank fields \citep{ste96,bou08, bou10, bun10}.  Studies based on the dropout selection method at $z\sim10$ suggest that the decrease in the characteristic luminosity of galaxies ($M^*$) with redshift seen at $3 < z < 7$ continues to $z\sim10$ and beyond \citep{yan10, bou11, bou12, oes12, ell13}.

A promising technique for increasing the number of source detections at high redshift is the ``cosmic telescope'' method \citep{zwi37} in which a foreground cluster is used to gravitationally lens distant faint objects into detectability.  Galaxy clusters can assist in constraining the luminosity functionby increasing the number of photons detected from distant sources, which would otherwise be too faint, and potentially boosting survey number counts \citep{ric06, ric08, sta07, bou09, bra09, hal12}.  Lensing probes a wider range of intrinsic luminosities, and when combined with blank field surveys targeting the bright end of the luminosity function, may even lead to a measurement of the faint end slope at $z > 7$ \citep{ric08}.  In addition to increasing number counts in deep surveys, lensing magnification makes more photons in individual sources available for spectroscopy \citep[e.g.,][]{bradac12}, helpful for following up detections made with photometric methods.   Lensing also provides improved spatial resolution, permitting the study of individual objects at intrinsic resolutions better than available today with single-aperture telescopes \citep[e.g.,][]{kne04, pel04, sch05, ric06, ric08, sta07, bra08, bra12, zhe09, zhe12, lap11, bou12, hal12, coe13}.  

Lensing is currently being used to constrain the $z \gtrsim 7$ luminosity function at fainter intrinsic magnitudes than blank field studies with comparable exposure time \citep{bro95b, ebb96, fra97, fry98, kne04, bra08, bou09, zhe09, bra12, hal12, zhe12, coe13}. In general, cosmic telescopes allow one to push 1-2 magnitudes fainter than the long-exposure observational limit of current telescopes, and so will be important for studies of the first stars and galaxies at $z > 10.$  However, the magnification also produces a loss of volume probed \citep{bro95a}, reducing detections of sources intrinsically brighter than the unlensed detection threshold.  

The usefulness of a cosmic telescope for increasing detected number counts can be quantified with the \'{e}tendue, or the areal coverage in the source plane with magnification exceeding a chosen threshold.  Cosmic telescopes with multiple projected structures along the line of sight can have larger \'{e}tendue than single-cluster lenses for two reasons.  First, lines of sight or ``beams'' with multiple projected structures can have higher total masses and thus larger regions of high magnification.  Second, multiple structures create interactions among the lensing planes that can boost the \'{e}tendue of the beam (\citeauthor{won12} \citeyear{won12}; see also \citeauthor{tor04} \citeyear{tor04}).  In this paper, we present J085007.6+360428 and J130657.5+463219 (hereafter referred to as ``0850'' and ``1306''), the first two lines of sight we have selected from the Sloan Digital Sky Survey \citep[SDSS;][]{ahn12} using Luminous Red Galaxies \citep[LRGs; e.g.,][]{eis01} to trace mass overdensities projected in $3\farcm5$-radius beams \citep{won13}.  The composition of the beams presented here reflect the diversity seen in the overall LRG-selected sample, with a single massive cluster dominating 0850 \citep[Zwicky 1953;][]{zwi61} and multiple clusters of up to $2  \times10^{15} M_{\odot}$ each populating 1306, one of which is Abell 1682 \citep{abe89}. 
  
In this paper we present the first spectroscopic survey of the line-of-sight structures in these beams out to $z \sim 0.8$.  We use MMT Hectospec spectroscopy to constrain the virial masses and radii of these structures in each beam.  We also present public Subaru Suprime-Cam \citep{miy02} imaging \citep{has08} that reveals several strongly lensed arcs visible in ground-based seeing, including a new candidate multiply-imaged source at $z=5.03$.  We test that the locations of critical curves predicted by the mass model derived from our spectroscopy and SDSS imaging alone are consistent with the coordinates of the detected arcs.  

Section \ref{sect:selection} summarizes the technique used by \citet{won13} to select our parent sample of high-mass beams from the SDSS, as well as what is known in the literature about beams 0850 and 1306.  Section \ref{sect:data} presents the MMT Hectospec spectroscopy and Subaru Suprime-Cam imaging and data reduction for both.  Section \ref{sect:results_discussion} discusses the spatial and kinematic properties of the structures.   Section \ref{sect:magnification_maps} describes the mass models and resulting magnification maps derived solely from spectroscopy and SDSS photometry.  We discuss uncertainties in those magnification maps.  Section \ref{sect:arc_analysis} describes the candidate multiply-imaged arcs discovered serendipitously in the Subaru imaging and compares them to the predicted critical curve locations.  Section \ref{sect:conclusions} summarizes our conclusions.  A $\Lambda$CDM cosmology is assumed throughout, with $H_0 = 71$ km s$^{-1}$ Mpc$^{-1}$, $\Omega_M = 0.27,$ and $\Omega_\Lambda = 0.73.$  For readability, we omit the $h$ convention with the understanding that masses and radii are in units of $h_{71}^{-1}$ and luminosities are in units of $h_{71}^{-2}$.

\section{Beam Selection and Properties}
\label{sect:selection}

\subsection{Selection Technique}

Our parent sample of dense beams is selected from the SDSS Data Release 9 \citep[DR9;][]{ahn12}.  The beams are selected to have large concentrations of LRGs to identify beams with a large total mass and possibly multiple projected structures.  The details of the beam selection are presented by \citet{won13} and briefly summarized here.  We compute the total LRG rest-frame $i^\prime$-band luminosity in random $3\farcm5$-radius apertures in the redshift range $0.1 < z < 0.7$.  We then sort the list of beams by total LRG luminosity and identify the top 200 best beams.  LRGs are biased tracers of the underlying mass distribution, so we expect that the fields with the greatest luminosity in LRGs are also the beams with the greatest total mass.  Other cluster selection techniques searching for concentrations of red galaxies have been presented, such as the Cluster Red Sequence method \citep{gla00} and MAXBCG \citep{koe07}.  The $3\farcm5$ radius is chosen to select beams with a large region of high magnification and to match the typical field size of ground-based, infrared, multi-object spectrographs for follow-up.  The result of the selection is a sample of beams with total LRG luminosities of $1.4 \times 10^{12} < L_{i} < 2.5 \times 10^{12} L_{\odot}$.  

The two beams presented in this paper were originally selected from an earlier version of our list of beams based on the \citet{pad07} LRG catalog, and they are both within the top 200 beams presented by \citet{won13}.  Although we present beam centers as given by \citet{won13}, the centers of our spectroscopic fields differ from the quoted beam centers by $\sim$15$^{\prime\prime}$ in 0850 and $\sim$107$^{\prime\prime}$ in 1306.  Our spectroscopic targeting extends to at least twice and at most four times the LRG selection radius ($r = 3\farcm5$), ensuring that we still cover the significant mass peaks in both fields.

\subsection{Properties of 0850 and 1306}
\label{sect:beam_properties}

The two beams we present are the first selected from the parent sample of 200 described above.  Their basic properties are in Table \ref{tab:tabulated_results_1}.  Both harbor known massive galaxy clusters:  Zwicky 1953 for 0850 \citep[$z = 0.378$;][]{zwi61} and Abell 1682 for 1306 \citep[$z = 0.234$;][]{abe89}.  There are also two Gaussian Mixture Brightest Cluster Galaxy \citep[GMBCG;][]{hao10} associations within $3\farcm5$ of the field center of 0850:   J132.49437+36.10756 at a photometric redshift of $z=0.284$ and J132.52774+36.01979 at a photometric redshift of $z=0.241$.  There are two additional associations within $3\farcm5$ of the field center of 0850 extracted from SDSS \citep{wen09,wen12}:  WHL J085005.0+360409 at a photometric redshift of $z=0.364$ and WHL J084956.9+360333 at a photometric redshift of $z=0.453$.  

There are four additional associations within $3\farcm5$ of the field center of 1306, including WHL J130650.0+463333 at a redshift of $z=0.225$ \citep{wen12}, NSC J130639+463208 with a photometric redshift of $z=0.2508$ \citep{gal03}, GMBCG J196.70832+46.55927 at $z=0.245, $ and GMBCG J196.75262+46.56389 at $z=0.337$ \citep{hao10}.  We discuss the overlap between these photometrically-identified groups and the groups that we identify with galaxy spectroscopy in Section \ref{sect:halo_properties}. 	

0850 is a member of the ROSAT Brightest Cluster Catalog \citep{ebe98, cra99} and the Northern ROSAT All-Sky (NORAS) Galaxy Cluster Survey \citep{boh00}.  The most obvious cluster within 0850 --- Zwicky 1953 --- was identified optically in the Second Palomar Observatory Sky Survey \citep{djo99} with a photometric redshift of 0.314 \citep{gal03}.   The X-ray temperature for Zwicky 1953 within 0850 is $\langle kT\rangle = 7.37$ keV from Chandra \citep{cav09} and $14.5$ keV from ROSAT \citep{ebe98}.  \citet{cav08} gives a bolometric X-ray luminosity of $1.7 \times 10^{45}$ ergs s$^{-1}$.  \citet{mau08} gives an X-ray luminosity of $1.6 \times 10^{45}$ ergs s$^{-1}$ and an X-ray temperature of 7.3 keV.  \citet{rei11} gives an X-ray temperature of $7.6^{+0.5}_{-0.5}$ keV and an X-ray luminosity of $2.6^{+0.04}_{-0.04} \times 10^{45}$ ergs s$^{-1}$.  

The principal cluster in beam 1306 has been identified as Abell 1682 at redshift $z = 0.2339$ \citep{abe89}.  The X-ray temperature from ROSAT for Abell 1682 is $8.9$ keV \citep{ebe98}.   \citet{cav08} gives a bolometric X-ray luminosity of $7.9 \times 10^{44}$ ergs s$^{-1}$ for Abell 1682.  \citet{rei11} gives an X-ray temperature of $7.0^{+2.0}_{-2.0}$ keV and an X-ray luminosity of $1.5^{+0.17}_{-0.17} \times 10^{45}$ ergs s$^{-1}$.

The X-ray temperatures and luminosities listed above for 0850 and 1306 are substantial, indicating the presence of massive clusters.  Note that the X-ray luminosities and temperatures for these two beams found in the literature are representative of the lowest redshift, dominant cluster in each beam, and do not take into account the potentially larger contributions from line-of-sight mass at higher redshifts.

\begin{table*}
\caption{Beam Properties.  }
\begin{center}
\leavevmode
\begin{tabular}[t]{ccccccc}
\tableline\tableline
\multicolumn{1}{c}{Beam name} &\multicolumn{1}{c}{RA (J2000)} &\multicolumn{1}{c}{Dec (J2000)} & \multicolumn{1}{c}{$L_X$ }& \multicolumn{1}{c}{$L_{LRG} $}& \multicolumn{1}{c}{Number of LRGs} &\multicolumn{1}{c}{Most Massive Cluster} \\
& & & (erg s$^{-1}$) & ($L_{\odot}$) &  & \\
\tableline	\\
0850 & 08 50 07.92 & +36 04 13.7 & $3.4 \times 10^{44}$&  $8.9  \times 10^{11}$ & 12 & Zwicky 1953 (08 50 11.2 +36 04 21, $z = 0.378$)\\
1306 & 13 06 54.63 & +46 30 36.7 & $1.1 \times 10^{44}$ &  $7.4 \times 10^{11}$ & 11 & Abell 1682  (13 06 49.7 +46 32 59, $z = 0.234$)\\ 
\tableline\tableline
  
\tableline
\end{tabular}

\label{tab:tabulated_results_1}

\end{center}
{\bf Notes.}  Beam locations are given as the coordinates of the LRG on which the $3\farcm5$ radius circle used to count LRGs is centered.  Total LRG luminosities, numbers of LRGs, and beam centroids are taken from \citet{won13}.  X-ray luminosities are from \citet{ebe98}.  X-ray luminosities are in the 0.1-2.4 keV band, expressed in the rest frame of the most massive cluster.  References for cluster redshifts are \citet{zwi61} and \citet{abe89}.
\end{table*}

\section{Data and Reduction}

\label{sect:data}
\subsection{Galaxy Spectroscopy}

We have completed a redshift survey of $1151$ field galaxies in these two beams with the 6.5 m MMT telescope on Mt. Hopkins, Arizona, and the Hectospec, a multi-object optical spectrograph with 300 fibers accessing a 1 degree field of view \citep{fab05, min07}.  Fiber crowding limits the number of galaxies that can be targeted in the central $7\farcm0$ diameter of our fields to $\sim30-40$ per exposure.  However, many fibers are available beyond this diameter to sample the outskirts of the centralized structures.  

Hectospec's grating with 270 grooves mm$^{-1}$ delivers a spectral resolution of $\sim4.5-5.2\;$\AA~($R\sim750-1800$) and a spectral coverage of $3650-9200\;$\AA~with a central wavelength of $6500\;$\AA.  This wavelength coverage is selected to include [OII] $\lambda 3726$ and $\lambda 3729$ emission lines, Ca II H $\lambda 3968$ and K $\lambda 3934$ absorption, and H$\beta$ lines for galaxies in the range $0 < z < 0.7.$  This redshift range samples the structures associated with the LRGs in our catalog \citep{won13}.
 
Galaxies are selected for Hectospec spectroscopic follow-up from the SDSS DR9 photometric catalog using only magnitude cuts and SDSS' morphological star/galaxy discriminator.  Specialized ``bright'' configurations with one hour total exposure time include galaxies as identified by SDSS with $i_{AB} < 20.5$.  ``Faint'' configurations with 2 hour exposure times included SDSS targets with $20.5 < i_{AB} < 21.1.$  

Fiber configurations for Hectospec are designed with the CfA \emph{xfitfibs} software.  Configurations acquired before June 2011 typically include only targets within a 7 arcminute radius of the beam centers and configurations acquired afterwards include targets out to a 15 arcminute radius.  $5-10$ F stars are included in each fiber configuration to enable flux calibration and removal of atmospheric absorption features.  \emph{xfitfibs} also permits inclusion of fibers for lower priority objects that do not interfere with high-priority objects; when possible, we use this capability to reobserve fainter targets for which previous observations had not yielded a redshift.  $20-50$ sky fibers are distributed randomly in the $1^{\circ}$ Hectospec field, avoiding galaxies known to have SDSS spectroscopy, and at least $\sim5$ of these are forced to be in the central $3\farcm5$ radius.

\subsubsection{Data reduction}
We reduce Hectospec data using HSRED, a modification of the IDL SDSS pipeline written by R. Cool\footnote{http://www.astro.princeton.edu/$\sim$rcool/hsred/} \citep{pap06}. HSRED  computes a wavelength solution from HeNeAr arc lamp spectra, removes cosmic rays and flat-fields the 2-D images, extracts spectra using fiber traces determined from dome flat observations, and subtracts sky spectra averaged from sky fibers.

We determine the optimum number of sky fibers by comparing the noise induced by poor sky subtraction beyond $8000\;$\AA~in object fibers for sets of up to $200$ randomly-distributed sky fibers.   Increasing the number of sky fibers beyond $50$ results in no further discernible decrease in the noise.  We improve sky subtraction over that provided in the HSRED pipeline by adjusting the amplitude and wavelength of the sky spectrum for each fiber to minimize residuals about the [OI] $\lambda 5577\;$\AA, Na $5893\;$\AA, and [OI] $\lambda 6300\;$\AA~sky lines.  

We determine redshifts and object classifications for combined spectra with an automated code modified from the SDSS pipeline specBS for Hectospec \citep{pap06}.  The pipeline finds the best linear combination of template spectra to minimize $\chi^2$.  The template spectra include 6 types of galaxy spectra and 4 types of QSOs.  The galaxy templates range from early-type to late-type and include an LRG spectrum.  To prevent sky-subtraction residuals from biasing the redshift solutions, we set the inverse variances at the locations of the thirty most prominent sky lines in the optical region ($3800\;$\AA~$ < \lambda < 8000\;$\AA) to zero.  

We perform a visual inspection of all spectra and redshift solutions, assigning one of three classes:  `A' (redshift success), `B' (possible redshift success), and `C' (redshift failure).  Only `A' spectra are included in this paper, although the quantitative results do not change significantly if the `B' spectra are included in the analysis.  `A' redshifts are assigned to 90\% of the sample with reduced spectra.  We present 596 'A' spectra satisfying $0.01 < z < 1.0$ in beam 0850 and 555 'A' spectra satisfying $0.01 < z < 1.0$ in beam 1306.  The positions, heliocentric redshifts, magnitudes, and estimated errors are in Table \ref{tab:tabulated_results_2} for all galaxies with a secure, visually-confirmed assigned redshift.

\subsubsection{Velocity zeropoint}


The HSRED reduction package uses arc lamp spectra to determine the velocity zero point of the spectrograph on a nightly basis.  However, telescope and instrument flexure potentially change the velocity zero point between observations.  We assess the magnitude of this effect by measuring the velocities of prominent night sky lines in individual spectra.  For this procedure, we reduce each observation without subtracting the sky as measured from sky fibers and ignore the Heliocentric correction.  For all sky and object spectra in each observation, we fit a single Gaussian to three prominent night sky lines ($5577\;$\AA,~$5890\;$\AA,~$6300\;$\AA).  The average zeropoint offset for all observations is less than 10 km s$^{-1}$.  The dispersion of the zeropoint offset for individual observations varies from 4 to 7 km s$^{-1}$.  Because these zeropoint offsets are significantly smaller than the quoted random errors, and could be due at least in part to errors in centroiding night sky lines, we do not correct for them.

\subsubsection{Internal errors}

We assess internal redshift errors by comparing the redshifts measured for multiply-observed objects.  To assess agreement at bright magnitudes, we use 14 bright galaxies ($19 < i_{AB} < 20.5$) observed twice in one hour configurations.  For these objects, the mean velocity offset is $5.2$ km s$^{-1}$ and the standard deviation of the differences about this value is $\pm 34$ km s$^{-1}$.   We also use 42 re-observations of bright F stars ($15 < i_{AB} < 16$) to assess the internal error in very high signal-to-noise ratio cases.  For these stars, the mean velocity offset is $3.3 \pm 2.5$ km s$^{-1}$ and the dispersion is $\pm 16$ km s$^{-1}$.   No catastrophic failures ($|\Delta z| > 0.01$ for galaxies or $|\Delta V| > 100$ km s$^{-1}$ for stars) are seen in either of these samples.  

We also re-observe an entire ``faint'' configuration in poor seeing to estimate the redshift errors for galaxies with $20.5 < i_{AB} < 21.1$.  Both observations received a 2 hour exposure time.  The second observation yields an average signal-to-noise ratio that is 54\% of the first observation due to poor seeing.  Although this average signal-to-noise ratio is lower than all other configurations and is unrepresentative of the overall sample, we assess the agreement between re-observations to place a conservative upper limit on the incidence rate of redshift failures.  100 objects are classified as galaxies (not QSOs or stars) with `A' quality in the first configuration and 53 matching objects from this set are classified as galaxies with `A' quality in the second observation.  The reduced number of objects with `A' quality in the second observation is due to the low signal-to-noise ratio in those data.  All of the matching redshifts agree to within 0.1\% ($\sim 300$ km s$^{-1}$) with a maximum deviation of $252$ km s$^{-1}$.  For these matches, the mean velocity offset is $3.0 \pm 9.0$ km s$^{-1}$ and the dispersion is $\pm 63$ km s$^{-1}$.  The deviations are Gaussian-distributed with extended tails.  No matches with `A' quality are catastrophic failures, including 10 additional matches classified as stars or QSOs.   Because no catastrophic failures are seen for multiply-observed objects in bright configurations, the catastrophic failure rate is likely negligibly small ($< 1$\%).  Since the second observation of the ``faint'' configuration has a poor signal-to-noise ratio not representative of the other Hectospec configurations, this upper limit on the failure rate is likely to be a conservative estimate.

\subsubsection{Comparison to redshifts from external surveys}

In general, we select spectroscopic targets to avoid objects for which redshifts had been measured by other surveys.  However, 24 galaxies from our overall Hectospec sample with $17.5 < i < 19.6$ also have SDSS spectra \citep{aih11}.  The average difference between the Hectospec velocities and the SDSS velocities (both Heliocentric) is $-25.0 \pm 10$ km s$^{-1},$ using the standard error on the mean as $1\sigma$ confidence intervals on this measurement.  The sign of the difference is such that the average Hectospec redshift is blueshifted with respect to the average SDSS redshift.  The $1\sigma$ standard deviation about this value is $47.5$ km s$^{-1}$.   The distribution of differences has no extreme outliers, with minimum and maximum values of -101 and 60 km s$^{-1}$, respectively.  The systematic offset of $-25.0 \pm 10$ km s$^{-1}$ is less than the typical velocity error for objects with $18 < i < 21$ as measured with configuration re-observations, as described in the previous section, so we do not correct the zeropoint for this value.

\begin{table}

\begin{center}
\caption{Spectroscopic Data (Excerpt)}
\begin{tabular}[t]{cccc}
\tableline\tableline
\multicolumn{1}{c}{RA (J2000)} &\multicolumn{1}{c}{Dec (J2000)} &\multicolumn{1}{c}{Redshift} &\multicolumn{1}{c}{$i_{AB}$} \\
\tableline	\\
 08 50 34.04     & +36 00 55.5      & 0.1082  $\pm$ 0.00024 & 20.3 \\                                 
   08 50 13.17     & +36 01 45.1   &    0.3809 $\pm$  0.00013 & 19.0   \\
     08 49 58.66    & +36 03 57.7    &  0.3753  $\pm$  0.00019 & 20.1  \\
        08 50 04.93   & +36 04 11.8    &   0.3654 $\pm$ 0.00015 & 19.3  \\
            08 50 21.88   & +36 03 23.3   &    0.6362 $\pm$ 0.00015 &  20.0    \\
                08 50 28.51 & +36 03 16.7 & 0.2173 $\pm$ 0.00015 & 19.1 \\

\tableline\tableline
		  
\tableline
\end{tabular}

\label{tab:tabulated_results_2}

\end{center}

{\bf Notes.}  Table 2 is published in its entirety in the electronic version of this publication. 
\end{table}

\subsubsection{Completeness}

\begin{figure*}[h]

  \begin{center}

 \plotone{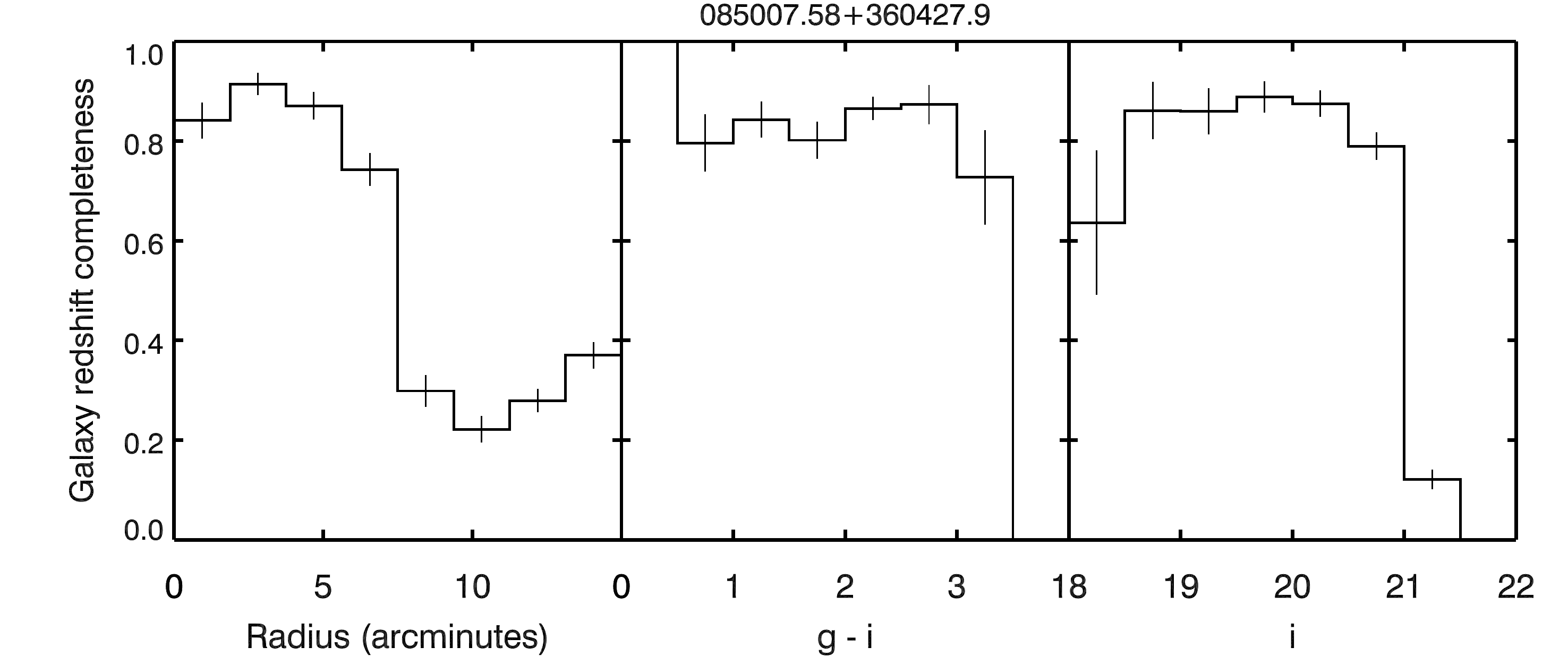}
 \plotone{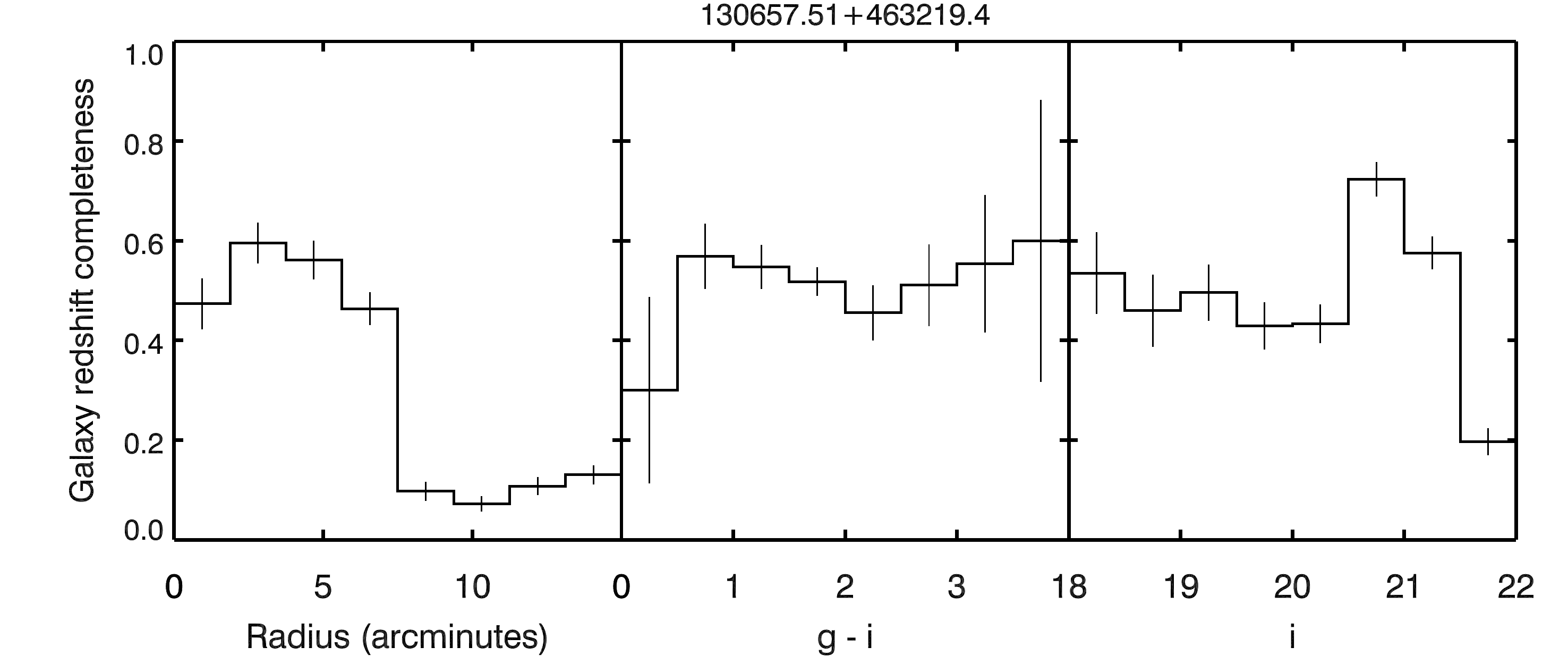}

  \caption{Redshift success completeness as a function of field radius, galaxy $g-i$ color, and galaxy $i$ magnitude for 596 galaxies in beam 0850 and 555 galaxies in 1306.  Error bars for individual bins are calculated from binomial statistics.  Within the error bars, the completeness is flat as a function of field radius, galaxy color, and magnitude, with the exception of the sparse sampling beyond $r = 7^{\prime}$ and at magnitudes fainter than $i_{AB} = 20.5$ in both beams.}
    \label{fig:completeness_radius}
 \end{center}
\end{figure*}

	\begin{table*}

\begin{center}
\caption{Imaging data.}

\begin{tabular}[t]{cccccc}
\tableline\tableline
\multicolumn{1}{c}{Beam name} & \multicolumn{1}{c}{Telescope/Instrument} &\multicolumn{1}{c}{Band}&\multicolumn{1}{c}{Date} &\multicolumn{1}{c}{Depth\footnote{Sensitivities are calculated from final stacked images using $1\farcs5$ diameter apertures.}} &\multicolumn{1}{c}{Exposure time}\\
&&&&  (3$\sigma$, AB) & (min) \\ 

\tableline	\\
0850 & Subaru/Suprime-Cam & B & 2006 Dec 20 & 27.3 & 44 \\
&Subaru/Suprime-Cam &  V & 2004 Feb 23, 2005 Nov 29 & 27.3 & 52 \\
& Subaru/Suprime-Cam & $R_c$ & 2000 Dec 26, 2005 Mar 4-5 & 27.5 & 70 \\
&Subaru/Suprime-Cam &  $I_c$ & 2000 Dec 26 & 26.7 & 56  \\
& Subaru/Suprime-Cam & i$^{\prime}$ & 2005 Mar 5 & 26.6 & 30  \\
&Subaru/Suprime-Cam &  z$^{\prime}$ & 2005 Mar 5, 2003 Apr 26 & 26.2 & 62  \\
&LBT/LUCI &  J & 2013 Mar 3 & 23.3 & 120  \\
1306 &Subaru/Suprime-Cam &  V & 2010 Mar 13-15 & 26.8 & 20  \\
&Subaru/Suprime-Cam &  i$^{\prime}$ & 2010 Mar 13-15 & 26.5 & 22  \\
\tableline\tableline
		  
\tableline
\end{tabular}



\label{tab:tabulated_results_3}
\end{center}
\end{table*}

Figure \ref{fig:completeness_radius} plots the redshift success completeness, or the ratio of the number of `A' quality spectra to the number of SDSS-identified galaxies.  We plot the completeness as a function of radius from the center of the field, galaxy color, and galaxy brightness.  The completeness in beam 0850 is high, exceeding 80\% for bright, central galaxies.  Some spectroscopic coverage beyond a radius of $r=7.5^{\prime}$ is available for both beams.  Within a radius of $7.5^{\prime}$, 0850's completeness is generally flat as a function of field radius, target color, and target brightness.  With completeness of 40\% for bright central galaxies, beam 1306 is less complete than 0850. although it has some coverage at very faint magnitudes ($i^{\prime} > 21.1$).  

\subsection{Archival Subaru Imaging}
\label{sect:subaru_imaging}
To search for lensed arcs in these beams, we make use of deep Subaru Suprime-Cam imaging in $B$, $V$, $R_c$, $I_c$, $i^{\prime}$, and $z^{\prime}$ bands.  These data are obtained from the Subaru-Mitaka-Okayama-Kiso Archive \citep[SMOKA;][]{bab02}.  Images taken for 0850 are obtained as part of the MACS \citep[Massive Cluster Survey; ][]{ebe01} followup program and have been published previously \citep{has08}.  Table \ref{tab:tabulated_results_3} summarizes the imaging depths, exposure times, and bands.

\label{sect:subaru_data_reduction}
Subaru Suprime-Cam images taken before January 2010 are reduced using the SDFRED1 package \citep{yag02, ouc04} and those afterward with the SDFRED2 package \citep{yag02, ouc04}.  We reduce images for 0850 with SDFRED1 and those for 1306 with SDFRED2.  Both reduction pipelines subtract overscan and bias frames, combine flat field frames, correct the frames for distortion and atmospheric dispersion, subtract the sky background, mask the AG shade and bad pixels, and finally align, scale, and combine the science frames.  Flat frames are constructed from science frames in all bands for both 0850 and 1306.  
 
Astrometric solutions and photometric zeropoints are determined for each band by matching Subaru star positions and photometry to sources detected in SDSS.  We first run the SDFRED1 program \emph{starselect.csh} on the Subaru bands, which calls Source Extractor \citep{ber96} to measure source fluxes and dimensions.  We select objects with peak fluxes less than 10,000 counts (unsaturated) lacking detected neighbors with more total flux than 2500 counts ($m \sim24-25$) within a 6$^{\prime\prime}$ radius (isolated) and a Kron radius less than 2.4 pixels (stellar).  The star positions are matched to SDSS coordinates with the program \emph{match-0.14}, which uses the method of similar triangles \citep{val95}.  We use a linear model matching shift, rotation, and plate scale.  Subaru stellar photometry is measured with 6$^{\prime\prime}$ diameter apertures with an aperture correction determined from more isolated stars (typically less than 0.03 magnitudes).  SDSS PSF magnitudes in the $u^{\prime}g^{\prime}r^{\prime}i^{\prime}z^{\prime}$ system are converted into UBVRI magnitudes with transformations on the SDSS DR5 website\footnote{http://www.sdss.org/dr5/algorithms/sdssUBVRITransform.html} credited to R. Lupton, with quoted errors typically less than 1\%.  

Zeropoints, photometric agreement, and astrometric agreement are assessed for 50-300 stars in each band.  Only objects classified as stellar by SDSS and satisfying the isolation and Kron radii criteria noted above are used for zeropoint calibrations.  After setting photometric zeropoints, the average agreement between Subaru aperture photometry and SDSS PSF magnitudes over all bands is 0.065 magnitudes ($1\sigma$ dispersion).  To assess the photometric noise attributable to Subaru photometry alone, we subtract the known SDSS photometric errors for each star in quadrature from the measured dispersion for each band, and find an average $1\sigma$ dispersion of 0.056 magnitudes.  The worst photometric agreement is seen in the $R_C$ band, with $0.101$ magnitudes of dispersion. The average dispersion in the astrometric positions with a linear model is 0.077 arcseconds.  

\subsection{LBT LUCI J-band Imaging}
\label{sect:luci_imaging}

We obtain 120 minutes of J-band imaging of 0850 with the LUCI infrared imager \citep{man08, age10} at the Large Binocular Telescope \citep[LBT;][]{hil08}.  Table \ref{tab:tabulated_results_3} summarizes the imaging depth and exposure time.  We reduce the data using simple IDL routines.  A master dark image is median combined from individual dark frames.  A master flat is median combined from individual dark-subtracted flat frames.  Science frames are dark subtracted and flat-fielded.  For these observations, LUCI exhibits repeatable pattern noise resembling sine waves; these are filtered out of the science frames by stacking rows of pixels with a 3-sigma clip perpendicular to the sine wave variation and subtracting the trend for each amplifier.  We run Source Extractor on each frame to measure source fluxes and astrometry, and then match the stars to seven known 2MASS sources with $H < 15$ in the field with \emph{match-0.14}.  For each frame, we adjust shift and rotation according to the \emph{match-0.14} results before stacking with a 3-sigma clip. 

\section{Results and Discussion}
\label{sect:results_discussion}

In this section, we describe the construction of group catalogs for each of the beams.  Assessing halo membership and estimating velocity dispersions, virial radii, and virial masses reliably are essential to modeling the field magnification.  Our approach for assigning galaxy membership and computing velocity dispersions follows techniques in the papers by \citet{zab90}, \citet{dan80}, \citet{gir98}, and \citet{biv06}.  The advantage of this approach is that masses can be estimated dynamically from galaxy redshifts and positions alone.  The calculation can be performed for groups and sparsely-sampled clusters with as few as 10 members.  Virial masses are computed following \citet{gir98} with the assumption that the virial theorem holds.  While it is hard to test for virialization, we note that LRGs trace the densest regions, so that our LRG-filled halos are among the most likely to be virialized.

\subsection{Constructing Group Catalogs}	

To identify structures in each beam, we first visually search for peaks in a redshift histogram that includes all secure, visually-confirmed redshifts in the field.  Candidate halos are identified by selecting groups of more than 10 objects clustered over a cosmologically corrected velocity range of at most $1500$ km s$^{-1}$.   Peaks adjacent in redshift are considered part of the same parent halo if they are separated by less than $1500$ km s$^{-1}$.

For each candidate peak, galaxies are considered members of the halo if the velocity separation to the nearest member galaxy is less than $1500$ km s$^{-1}$.  Membership is assessed starting at the peak redshift and extending both blueward and redward.  For the resulting cluster members, we compute the mean cluster redshift $\overline{z}_{group}$ and projected velocity dispersion $\sigma_{los}$ using the biweight estimators \citep{bee90, mom09}.  Interlopers are trimmed from halo membership by removing galaxies separated by more than $1\sigma$ from member galaxies.

\subsection{Estimating Virial Radii and Velocity Dispersions}
\label{sect:virial_radii}

Quantities such as velocity dispersion and virial radius $R_{vir}$ are ideally estimated using galaxies within the virial radius, but the sensitive dependence of calculated virial radius on the aperture necessitates an iterative approach \citep{gir98}.  At the start of the iterations, we would like to apply an aperture $a$ with constant physical dimensions to each halo to avoid biasing halo properties with redshift.  This initial aperture must avoid extending beyond the field of view with spectroscopic coverage for all halo candidates.  

We calculate this initial aperture by first computing the peak centroid of all candidate peaks.  For each peak, we calculate the angular distance between the peak centroid and the edge of the field of view with spectroscopic coverage and convert this distance to physical units (Mpc).  We then find the minimum distance to the edge of the field in physical units over all the halos and set the aperture radius $a$ equal to this value.  For 0850 and 1306, these distances are 2.42 Mpc and 1.46 Mpc, respectively.  We perform the following steps for each candidate peak to arrive at self-consistent estimates of virial radius and velocity dispersion:

\begin{enumerate}
\item Within the aperture $a$, calculate mean velocity and and velocity dispersion using the biweight estimators.
\item Restrict membership in redshift space using a conservative $3\sigma$ cut with respect to the peak's central velocity.  Membership is assigned to galaxies satisfying \citep{zab90, yah97}:
\begin{equation}
c \delta_z < 3 \sigma_{in}(1+z)
\end{equation}
\item Within the aperture $a$, recalculate peak centroid, mean velocity, and velocity dispersion with biweight estimators.
\item Estimate the virial radius with equation (9) of \citet{gir98}:
\begin{equation}
R_{vir} = 0.002 \sigma
\end{equation}
where $\sigma$ is in units of km s$^{-1}$ and $R_{vir}$ is in units of Mpc.  Set the aperture $a=R_{vir}$.
\item Restrict membership with a $3\sigma$ cut in velocity and recalculate peak centroid, mean velocity, and velocity dispersion within aperture $a$ with biweight estimators.
\item Calculate the projected harmonic mean radius within the aperture $a$ with equation (7) of \citet{gir98}:
\begin{equation}
R_{PV} = \frac{N(N-1)}{\sum_{i>j}R_{ij}^{-1}}
\end{equation}
where $R_{ij}$ is the distance between the $i$th and $j$th members and $N$ is the number of halo members.
\item Estimate the virial radius with a modified version of equation (9) from \citet{gir98}:
\begin{equation}
R_{vir}^3 = \frac{3 \pi \sigma^2 R_{PV}}{\Delta_h H_0^2 \Omega_M  (1+z)^3},
\end{equation}
where the factor $(1+z)^3$ includes the dependency of the background density on redshift.   The parameter $\Delta_h$ is defined as the ratio of the virial density to the background density at the redshift of formation, $\rho_{vir} = \Delta_h \overline{\rho}(z)$.  Conventions for $\Delta_h$ found in the literature include 180, 200, $200 / \Omega_{M} (z),$ etc.  Following \citet{zha09}, we adopt a functional form for $\Delta_{h}$ presented by \citet{bry98} and derived from hydrodynamic simulations:
\begin{equation}
\Delta_h = \frac{18 \pi^2 + 82 x - 39 x^2}{\Omega_M},
\end{equation}
where $x = \Omega(z) - 1$,
\begin{equation}
\Omega(z) = \frac{\Omega_M (1+z)^3}{E(z)^2}, and
\end{equation}
\begin{equation}
E(z)^2 = (\Omega_M (1+z)^3 + \Omega_\Lambda).
\end{equation}
We use the peak redshift in the equations for the virial radius.  Following calculation of $R_{vir}$, we set the aperture $a$ equal to $R_{vir}$.
\end{enumerate}

We iterate steps 5-7 until the procedure has converged or the number of members is zero.  Halo membership is re-assessed during each iteration.  We define convergence as the point at which the number of members and velocity dispersion estimate no longer changes from the previous iteration to the next.  For all the halo candidates we identify in these two beams, the procedure converges or ends with zero members before 10 iterations are reached.  Following the convergence, we estimate the virial mass of the halo with equation (5) from \citet{gir98}, originally from \citet{lim60}:
\begin{equation}
M_V = \frac{3\pi}{2}\frac{\sigma^2 R_{PV}}{G}
\end{equation}

Note that this formula depends on the projected harmonic mean radius and not on the virial radius.  The calculated halo masses, velocity dispersions, virial radii, and halo membership do not change appreciably when steps 2-4 are omitted, i.e., when the first estimate for virial radius is calculated directly from the projected harmonic mean radius rather than using equation (9) of \citet{gir98}.

\begin{figure*}[h]

  \begin{center}
    \epsscale{0.8}
 \plotone{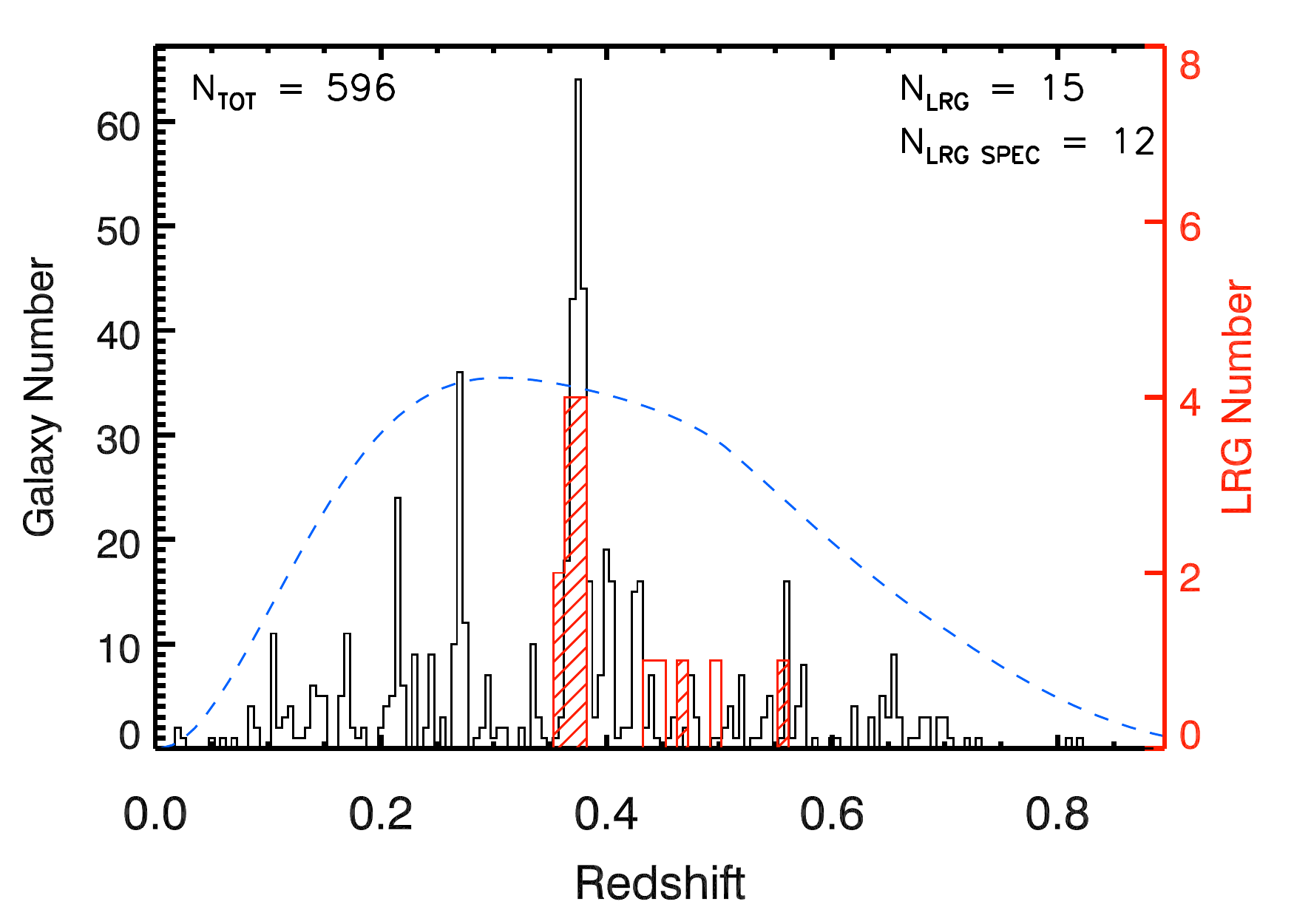}
  \caption{Galaxy redshift histogram for galaxies in beam 0850 for $30^{\prime}$ field diameter.  The bin size is 1500 km s$^{-1}$.  A redshift histogram for LRGs is overlaid in red, with photometric redshifts replaced by spectroscopic redshifts when available from SDSS or our survey.  Spectroscopic redshifts for LRGs are denoted by a shaded histogram.  The bin size for LRG histograms is 3000 km s$^{-1}$ and the LRG counts are marked on the rightmost y-axis.  The blue dashed line is the redshift selection function for a homogeneous universe calculated using redshift completeness as a function of magnitude and field size with published luminosity functions for $0 < z < 1$ \citep{fab07}.  The y-axis normalization for the redshift selection function is arbitrary.  K-corrections are calculated by transforming rest-frame $M_B$ into observed $i^\prime$ using an LRG spectral template following \citet{hog02}.  The histogram indicates the presence of multiple candidate structures in addition to a dominant structure at $z\sim0.375$.}
    \label{fig:0850_histogram}
 \end{center}
\end{figure*}

\begin{figure*}[h]

  \begin{center}
    \epsscale{0.8}
   \plotone{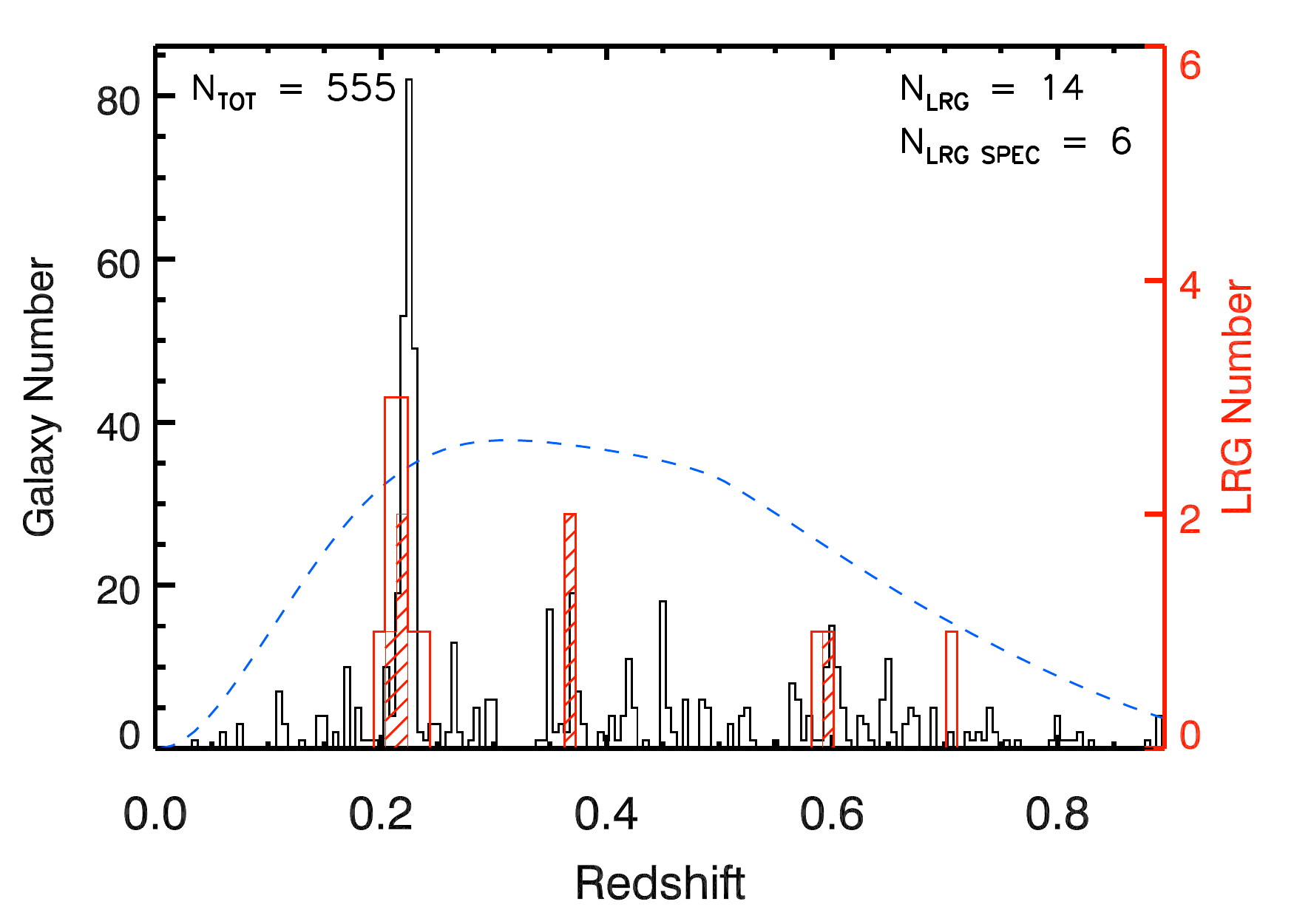}
  \caption{Galaxy redshift histogram for beam 1306 for $30^{\prime}$ field diameter.  Lines, symbols, and histograms follow Figure \ref{fig:0850_histogram}.  As in 0850, the histogram indicates the presence of multiple candidate structures in addition to a dominant structure at $z\sim0.22$.  There are three candidate structures flagged by LRGs with spectroscopic redshifts (shaded histograms), all of which correspond to peaks in the full galaxy redshift histogram.  Only one LRG with a photometric redshift (and without a spectroscopic redshift) fails to coincide with a peak in the full redshift histograms. }
  \label{fig:1306_histogram}
 \end{center}
\end{figure*}

\begin{figure*}[ht]

  \begin{center}
\epsscale{1.0}

 \plottwo{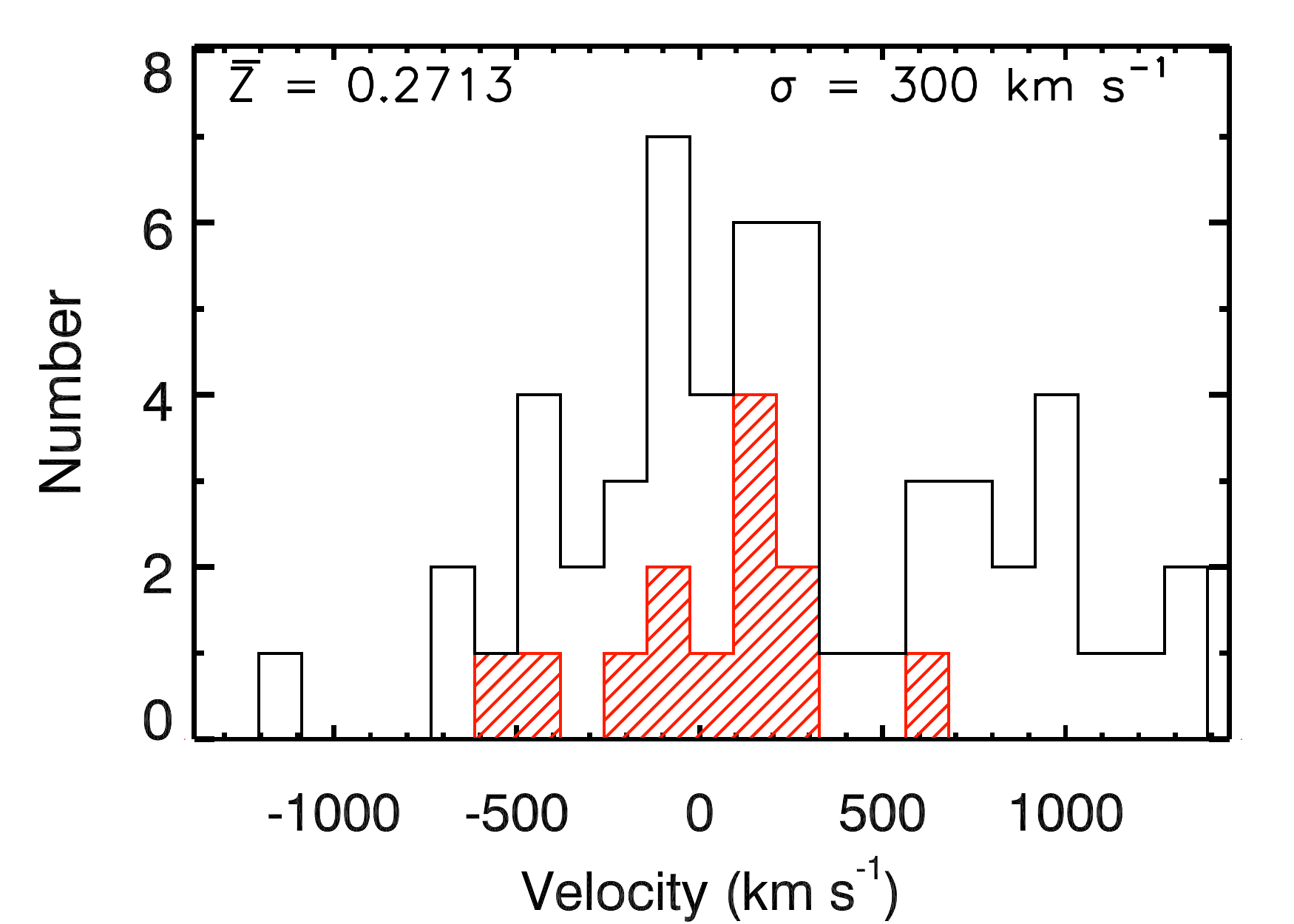}{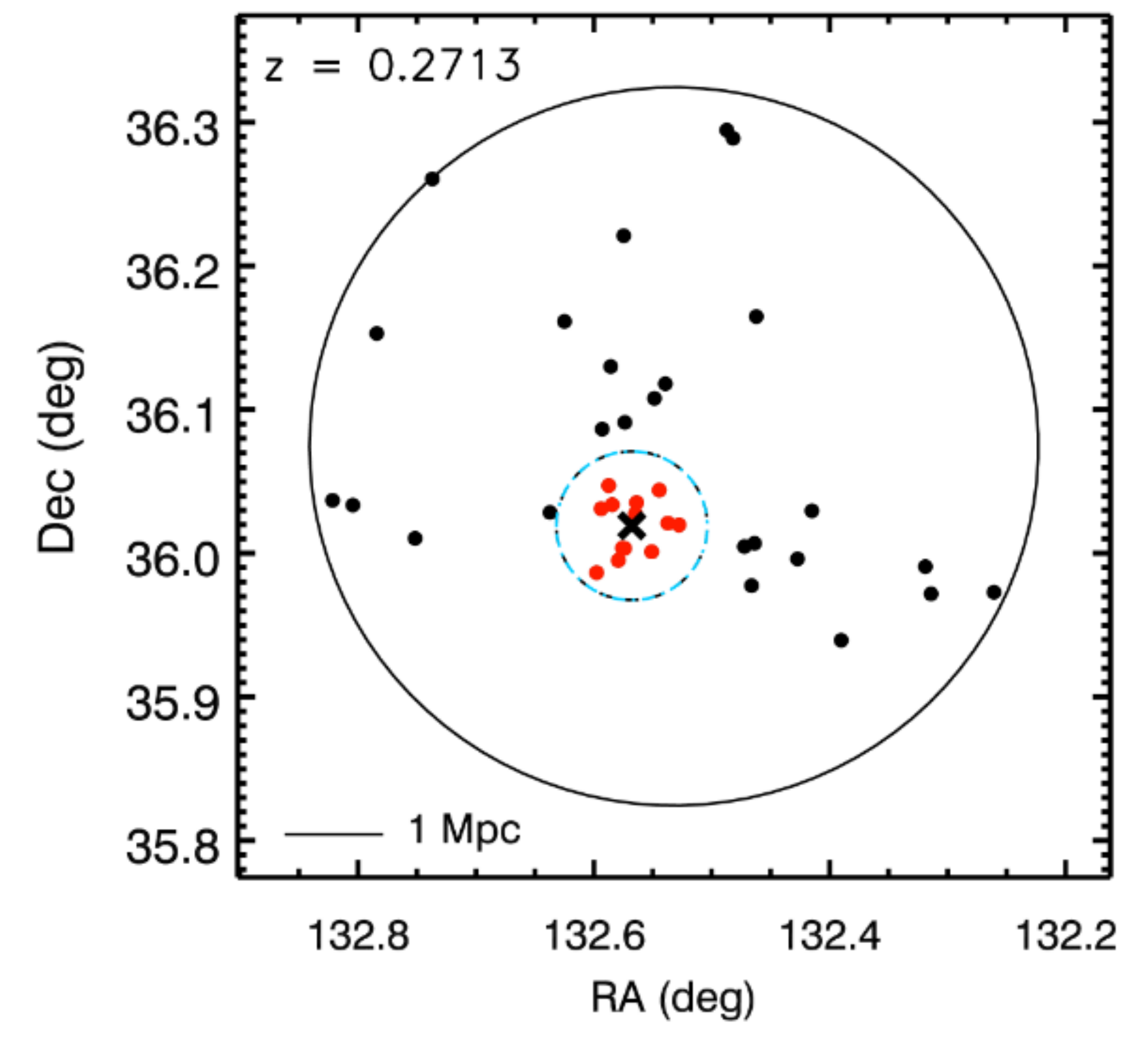}
  \plottwo{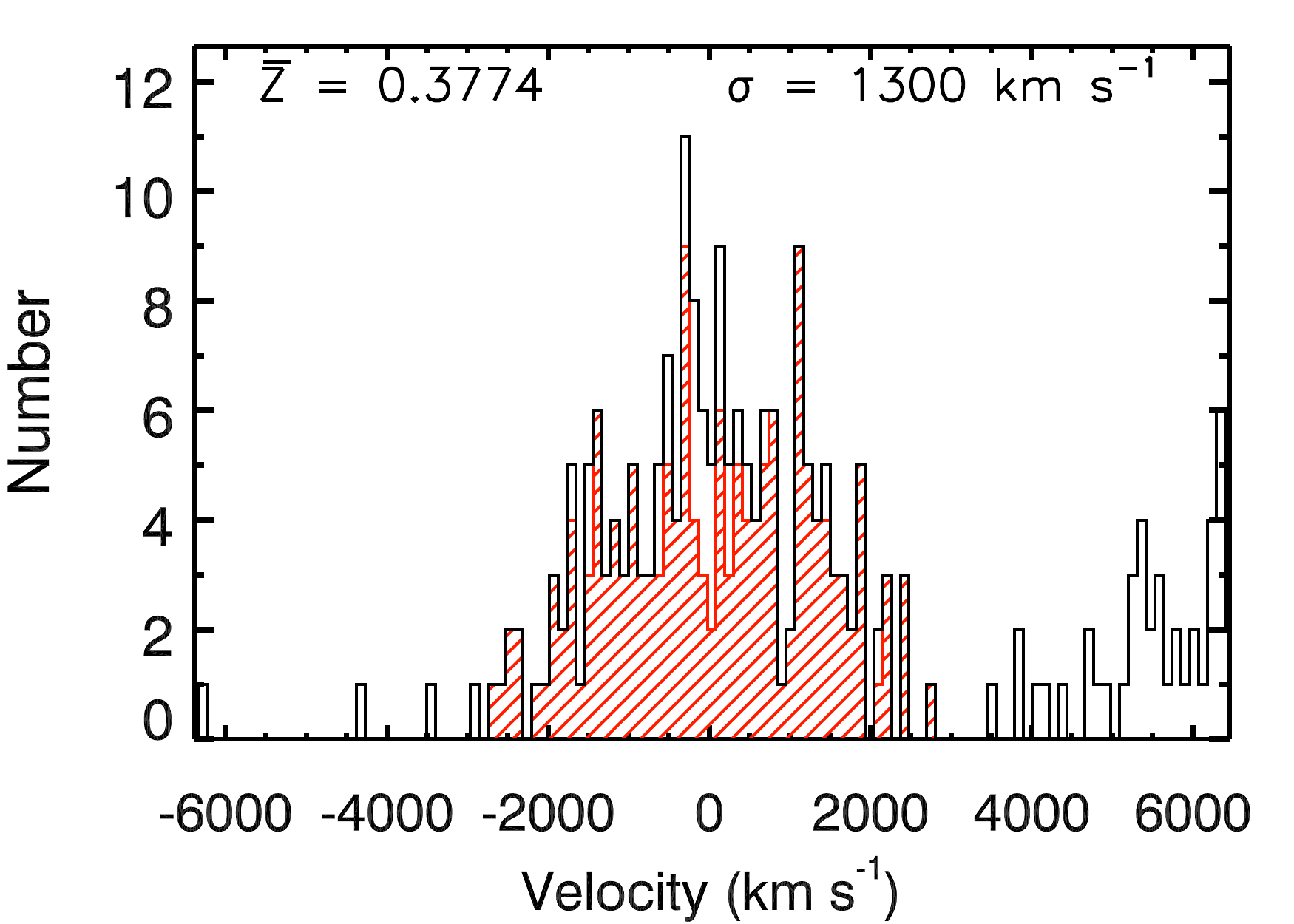}{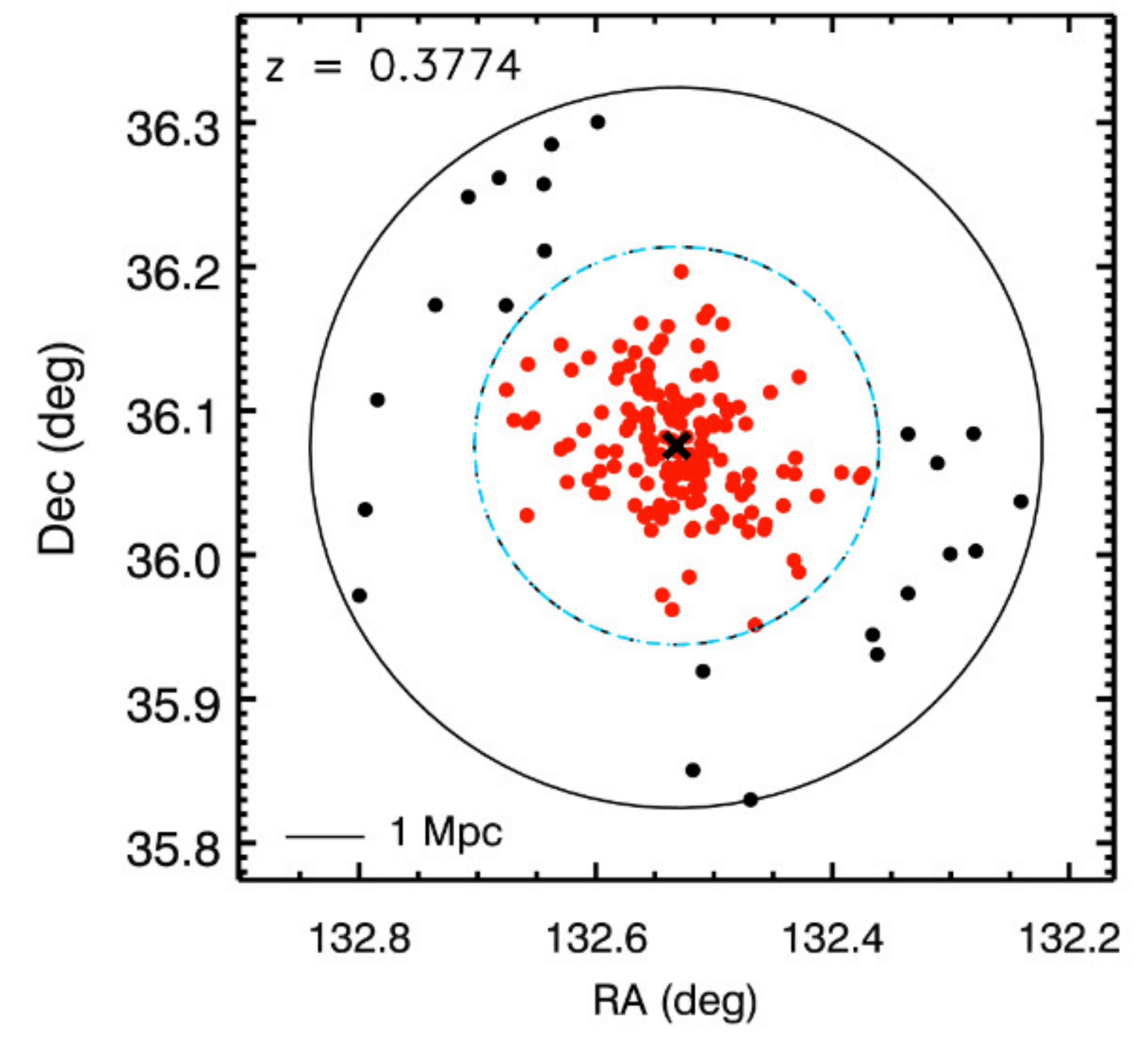}
  \caption{{\it Left:}  Galaxy velocity histograms for two surviving peaks in beam 0850.   The black histogram shows galaxies for a $30^{\prime}$ arcminute field diameter and the red shaded histogram shows final members as determined by the iterative procedure described in Section \ref{sect:virial_radii}.  Bin size is 150 km s$^{-1}$.  {\it Right:}  Field maps showing members and nonmembers in each peak.  The large black circle denotes the spectroscopic field, the black X marks the final peak centroid, and the blue dashed line shows the final calculated virial radius, which is used to select group members.  The red filled circles denote group members (corresponding to the red histogram in the left panels).  The black filled circles denote galaxies falling within $3\sigma$ of the peak velocity but are outside of the final calculated virial radius.  Note that only a minority of galaxies within $3\sigma$ of the peak velocity in 0850\_1 are classified as members, indicating substantial structure outside of a virial radius, although the velocity widths of  the member and non-member contributions are similar.  In both peaks, the virial radius is completely circumscribed by the field selection circle, suggesting that group membership is not affected by a lack of redshifts beyond $r = 15^{\prime}$.}
  \label{fig:0850_histogram_peaks}
 \end{center}
\end{figure*}

\begin{figure*}[h]

  \begin{center}
  \epsscale{1.0}
 \plottwo{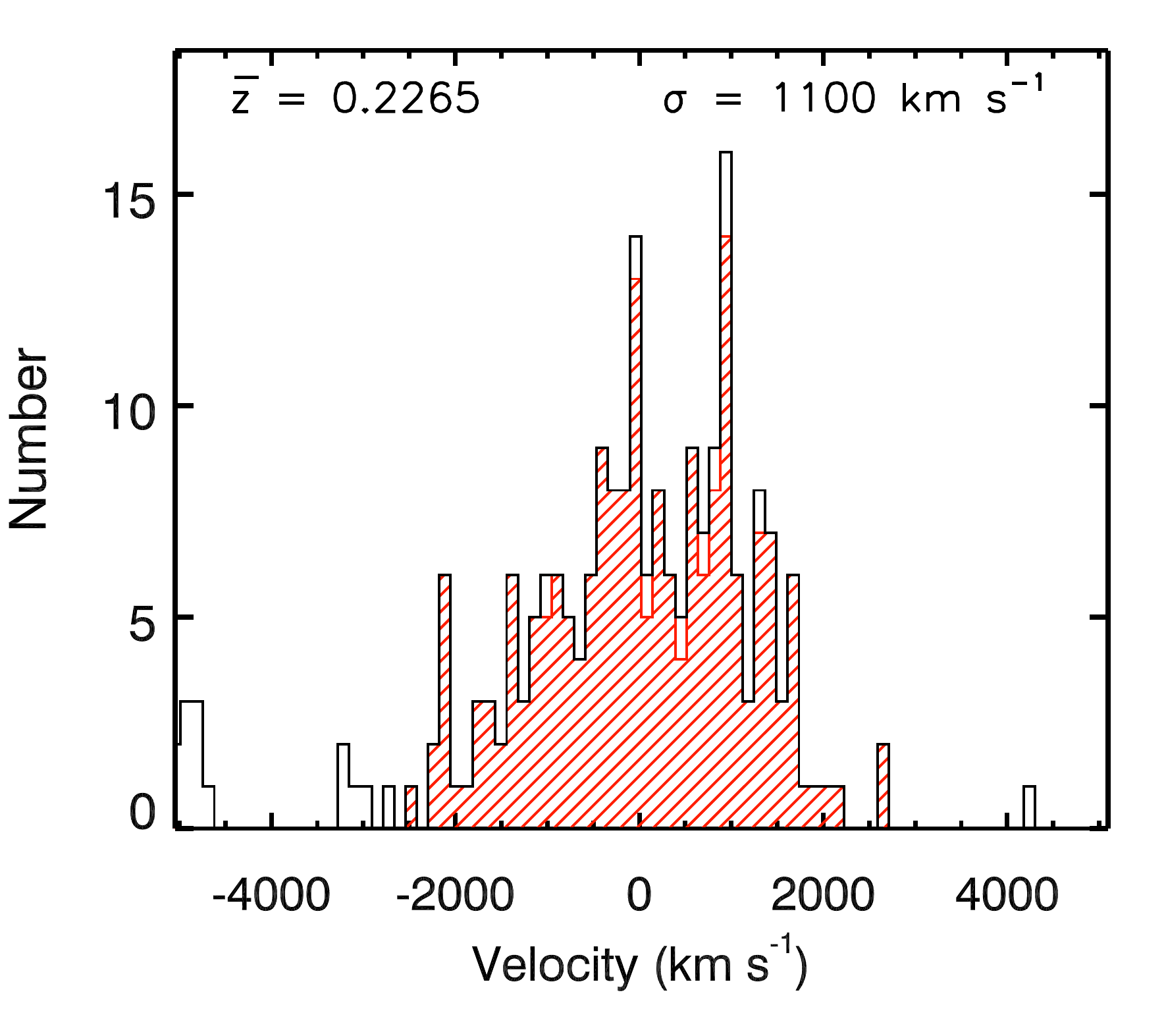}{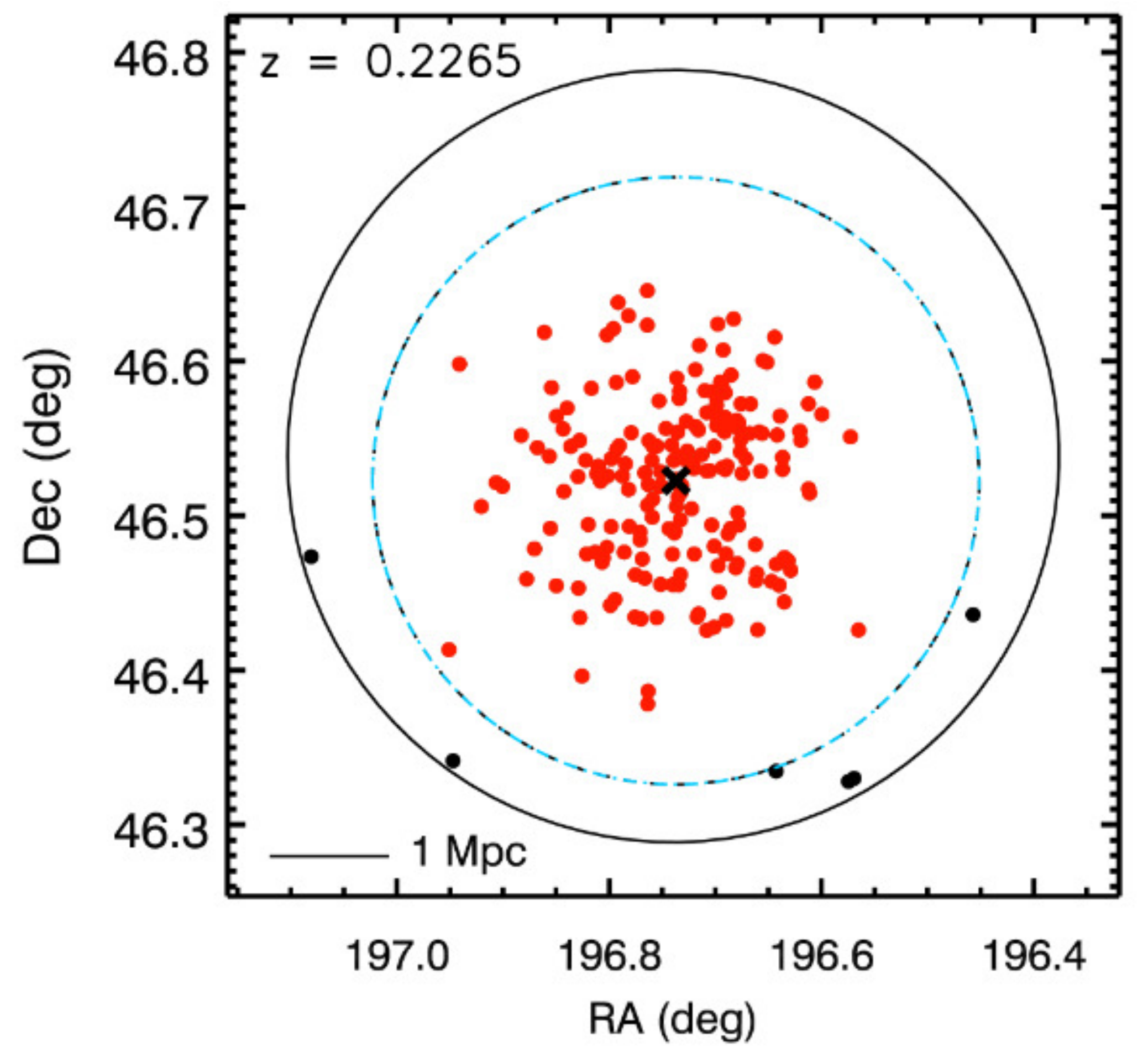}
 \plottwo{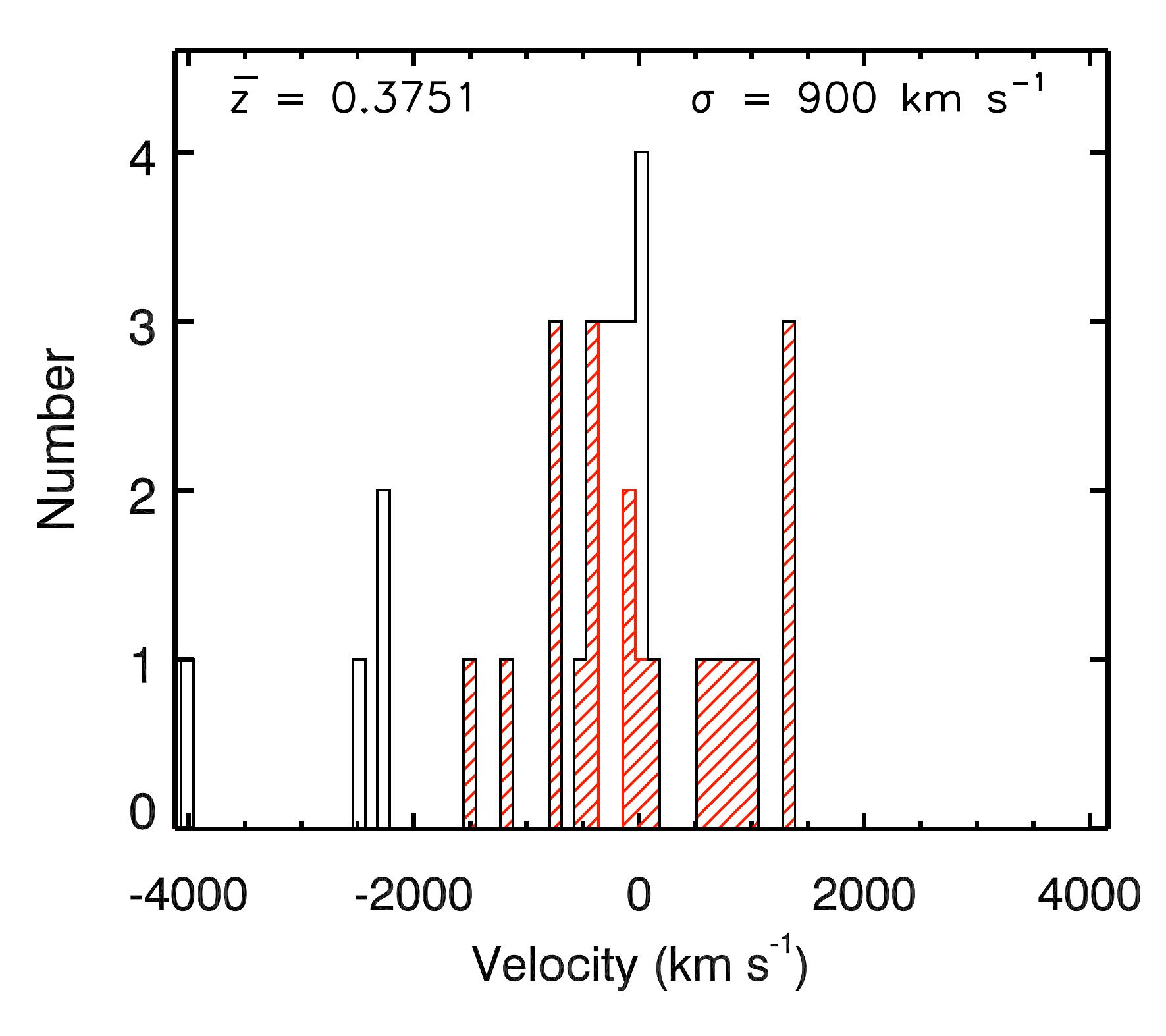}{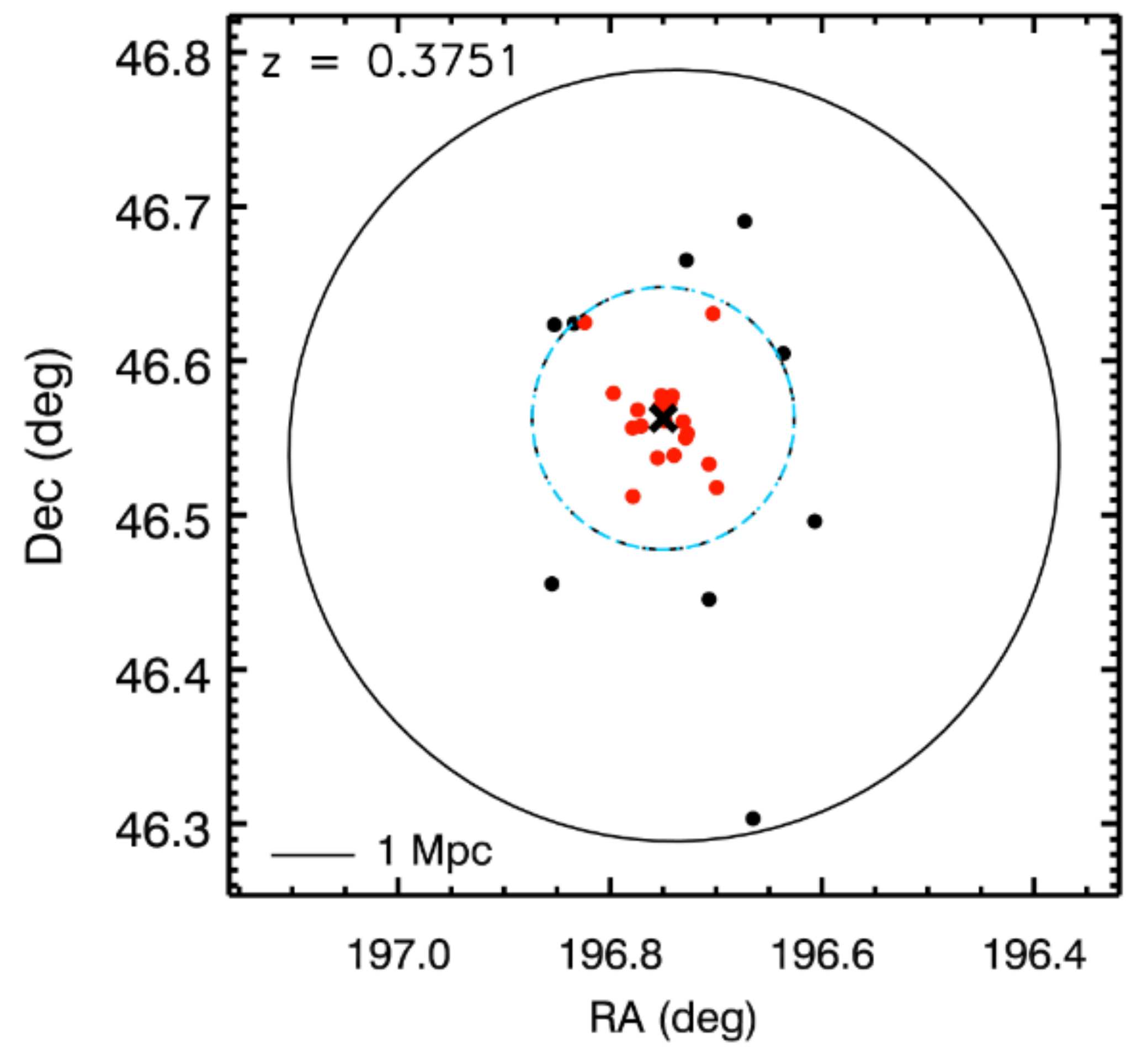}
 \plottwo{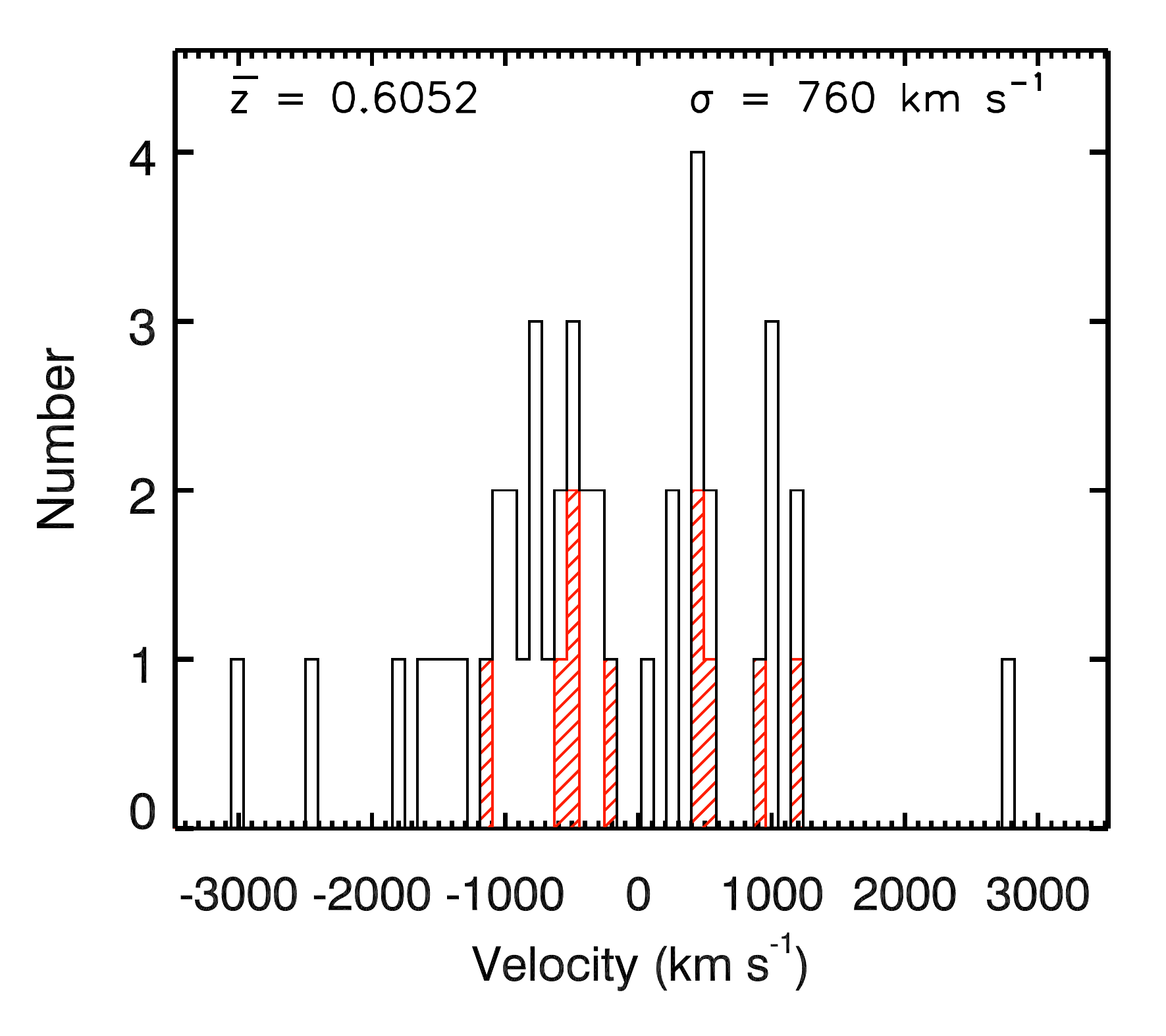}{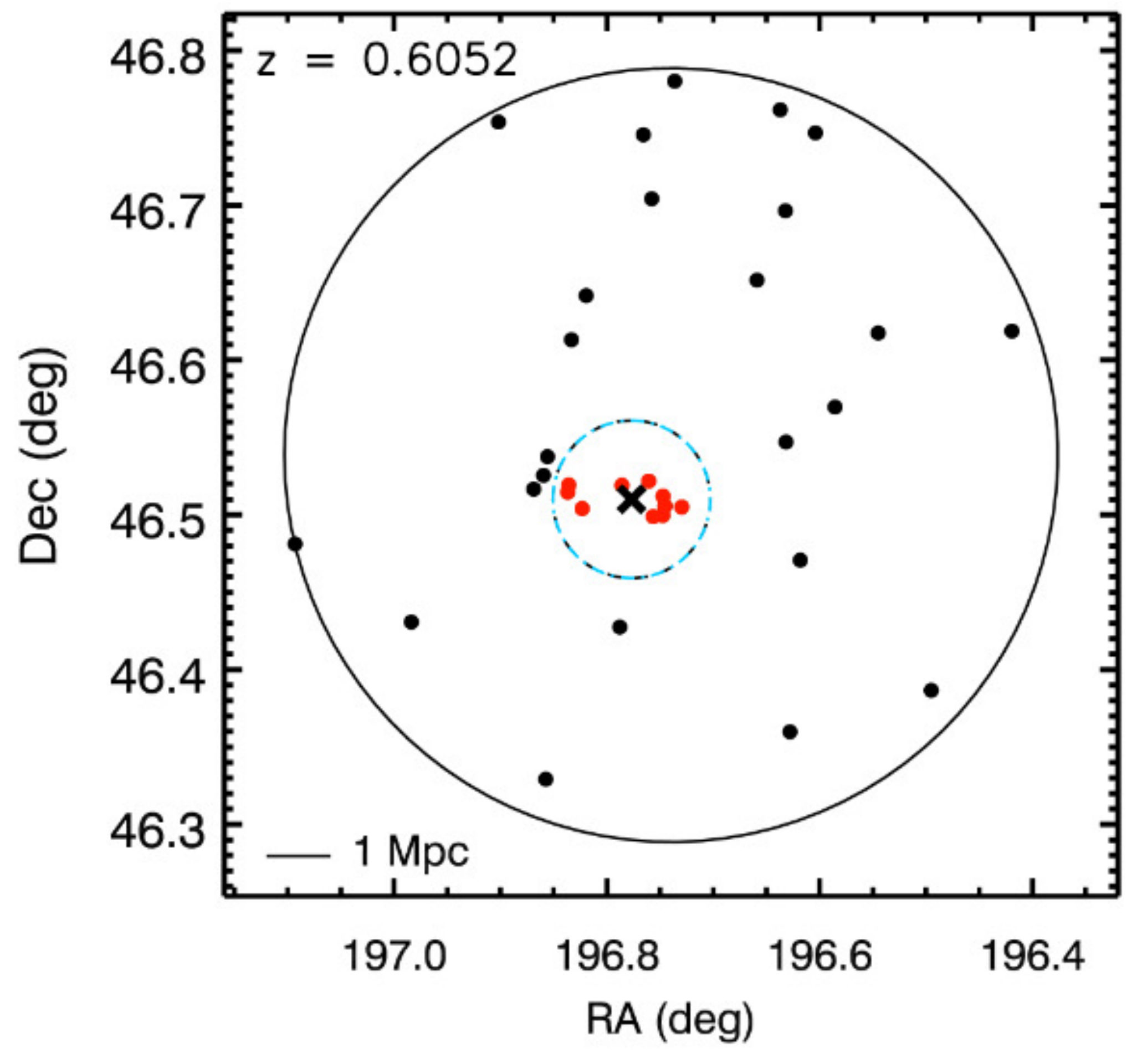}
  \caption{Galaxy velocity histograms and field maps for three peaks in beam 1306 for 30 arcminute field diameter.  Lines, histograms, and symbols follow Figure \ref{fig:0850_histogram_peaks}.  All peaks have membership circles that are circumscribable by the field radius, indicating that estimates of group properties are not affected by incompleteness outside of  $r=15^{\prime}$ for these peaks.}
  \label{fig:1306_histogram_peaks}
 \end{center}
\end{figure*}

\subsection{Properties of Identified Groups}
\label{sect:halo_properties}

\begin{table*}[h]
\caption{Group Candidates in Beams 0850 and 1306.}
\begin{center}
\leavevmode

\begin{tabular}[t]{cccccccc}
\tableline\tableline
\multicolumn{1}{c}{ID} &\multicolumn{1}{c}{$\overline{z}_{group}$\footnote{Velocity dispersion errors are determined with jackknife resampling of the logarithm of the velocities, using equation (22) in the paper by \citet{bee90}.  All other errors are determined by calculating 16-th and 84-th percentiles of parameter distributions using bootstrap resampling.  We use the jackknife of the logarithm of the velocities for calculating the error on the velocity dispersion here because that method outputs well-behaved errors for poorly sampled halos.  We employ the bootstrap errors for all tabulated quantities, including the velocity dispersion, in our analysis of the magnification maps throughout this paper. }  }  &\multicolumn{1}{c}{\# members}  &\multicolumn{1}{c}{Centroid}  &\multicolumn{1}{c}{Centroid Error } &\multicolumn{1}{c}{$\sigma_{los}$ } &\multicolumn{1}{c}{$M_{200} $} &\multicolumn{1}{c}{$R_v$}  \\
\multicolumn{1}{c}{} &\multicolumn{1}{c}{}    &\multicolumn{1}{c}{}  &\multicolumn{1}{c}{}  &\multicolumn{1}{c}{($^{\prime\prime}$)} &\multicolumn{1}{c}{(km s$^{-1}$)} &\multicolumn{1}{c}{$(\times 10^{14} \ensuremath{M_{\odot}}$)} &\multicolumn{1}{c}{(Mpc)}  \\
\tableline	\\
0850\_1 & $0.2713^{+0.0004}_{-0.0004} $ & 14 & 08 50 17.1 +36 01 13 & 22 & $300^{+110}_{-90}$& $0.6^{+0.4}_{-0.4}$ & $0.8^{+0.1}_{-0.2}$\\
0850\_2 & $0.3774^{+0.0004}_{-0.0005} $& 161 & 08 50 07.6 +36 04 35 & 12 & $1300^{+60}_{-60}$ & $32^{+3.1}_{-2.9}$ & $2.6^{+0.1}_{-0.1}$\\
1306\_1 & $0.2265^{+0.0003}_{-0.0003} $ & 195 & 13 06 56.1 +46 31 27 & 14 & $1100^{+50}_{-50}$& $20^{+1.9}_{-1.8}$ & $2.5^{+0.1}_{-0.1}$\\
1306\_2 & $0.3751^{+0.0008}_{-0.0009} $& 21 & 13 06 59.5 +46 33 31 & 18 & $900^{+130}_{-120} $&$7.4^{+1.8}_{-1.7}$ & $1.6^{+0.1}_{-0.1}$\\
1306\_3 & $0.6052^{+0.0011}_{-0.0012} $& 10 & 13 07 05.5 +46 30 36 & 30 & $760^{+150}_{-130} $ &$6.2^{+2.2}_{-2.0}$ & $1.2^{+0.1}_{-0.2}$\\
\tableline\tableline
		  
\tableline
\end{tabular}

\label{tab:tabulated_results_4}
\end{center}
\end{table*}

We treat groups that survive this iterative procedure with $\geq 10$ members as halos and include them in our mass model.  The properties of these structures are listed in Table \ref{tab:tabulated_results_4}.  We use the notation ``$0850\_1$'' to refer to separate halos within beams.  The velocity dispersion errors are determined with jackknife resampling of the logarithm of the velocities, as described by \citet{bee90}.  All of the other errors in Table \ref{tab:tabulated_results_4} are determined with bootstrap resampling.  Full redshift histograms for both beams are shown in Figures \ref{fig:0850_histogram} and \ref{fig:1306_histogram}.  Redshift histograms and field maps for each of the surviving groups are in Figures \ref{fig:0850_histogram_peaks} and \ref{fig:1306_histogram_peaks}.   

Many photometrically-identified LRGs \citep{won13} have Hectospec spectroscopic redshifts.  Those LRGs with spectroscopic redshifts within $r = 3\farcm5$ of the center of 0850 trace three structures, as shown in Figure \ref{fig:0850_histogram}; the mean redshifts of these LRGs are $z=0.3782$ with 10 LRGs, $z=0.4754$ with 1 LRG, and $z=0.5631$ with 1 LRG.  The first of these corresponds to cluster 0850\_2 identified with our full galaxy spectroscopy.  The other two are not associated with significant structures determined by our iterative procedure.  There is no candidate peak in the field spectroscopy near $z=0.4754$, suggesting that this LRG is either spurious or an isolated red galaxy in a small, sparsely sampled group.   There is a candidate peak at $z=0.563$ that may correspond to the third LRG at $z=0.561$.   That peak converges on a $\sigma=150 \pm 90 $ km s$^{-1}$ group with 4 members within the virial radius.  Since our field spectroscopy is not deep enough to recover this structure with substantial membership, we do not present this peak as a candidate structure; but given its very small velocity dispersion estimate, it is unlikely that it would contribute significantly to the field magnification if real.  

The LRGs with spectroscopic redshifts within $r = 3\farcm5$ of the center of 1306 mark three structures, as shown in Figure \ref{fig:1306_histogram}; the mean redshifts of these LRGS are $z=0.2233$ with 3 LRGs, $z=0.3741$ with 2 LRGs, and $z=0.6019$ with 1 LRG.  The redshifts of the three groups identified by our full spectroscopy (1306\_1, 1306\_2, and 1306\_3) agree with these three structures, suggesting that they are real.  The velocity dispersions of the two groups 1306\_2 and 1306\_3 are substantial ($900^{+110}_{-110} $ and $760^{+130}_{-150} $ km s$^{-1}$, respectively), yet these groups are sparsely sampled with 10-18 members, so their association with LRGs is important in interpreting them as real.

Overall, in both 0850 and 1306, 16 of 18 LRGs (89\%) with spectroscopic redshifts are associated with halos that survive our iterative procedure, and 17 of 18 LRGs (94\%) are associated with at least a visible peak in the full redshift histograms.

As discussed in Section \ref{sect:beam_properties}, previous studies have identified associations in these beams, including Zwicky 1953 in 0850 and Abell 1682 in 1306.  Zwicky 1953 in 0850 at $z=0.378$ \citep{zwi61} is very likely the same as our massive structure 0850\_2 at $z=0.3774$.  0850 also has two photometrically identified associations at $z=0.241$ and $z=0.284$ \citep{hao10} within $3\farcm5$.  Since the redshifts of these associations are generated solely from photometric data, cross-identifications with our spectroscopically-derived catalogs must be approached cautiously.  It is not immediately clear how these associations relate to the two groups we identify in the spectroscopy, although it is plausible that the $z=0.284$ structure corresponds to Zwicky 1953 and the lower redshift object corresponds to the small group we find at $z=0.2715.$  \citet{wen12} find two associations photometrically at $z=0.364$ and $z=0.453$; it is clear that the first of these corresponds to the massive cluster 0850\_2 we identify at $z=0.3774$.  Although the second of these does not appear to be associated with a peak in the redshift histogram, its photometric redshift approximately matches the spectroscopic redshift of an LRG at $z=0.4754$.

Abell 1682 in 1306 at $z=0.2339$ \citep{abe89} corresponds to the massive structure 1306\_1 we find at $z=0.2265.$  There are four other photometrically identified associations in the field of 1306, three of which are nearly the same redshift and likely correspond to Abell 1682:  $z=0.2081$ by \citet{wen09}, $z=0.245$ by \citet{hao10}, and $z=0.2508$ by \citet{gal03}.   The last photometric association at $z=0.337$ \citep{hao10} may correspond to the structure we identify at $z=0.3746.$

Abell 1682 has a published X-ray-derived mass of $M_{500} = 1.2 \times 10^{15} \Msun$ \citep{man10, rei11}.  To enable comparison to our spectroscopically-derived $M_{200}$ measurement, we calculate the expected $M_{200} / M_{500}$ ratio for an NFW halo.  We assume a concentration range consistent with the mass-concentration relation in the paper by \citet{duf08}, or 3 - 4 for a $\sim 1 \times 10^{15} \Msun$ halo at $z \sim 0$.  This results in a $M_{200} / M_{500}$ range of 1.44 - 1.54, giving an X-ray-derived virial mass of $M_{200, X-ray} = 1.8 - 1.9 \times 10^{15} \Msun$.  This value is in excellent agreement with our spectroscopically derived value of $M_{200, spec} = 2.0^{+0.19}_{-0.18}  \times 10^{15} \Msun$.

In general, the halo parameters for the most massive halo in 0850 are better constrained than several of the more massive halos in 1306.  This is largely due to incomplete spectroscopic sampling of two massive components in 1306.  The component 1306\_2 at $z = 0.3746$ has uncertainties of $\sim36\%$ on the mass, $\sim16\%$ on the virial radius, and a centroid error of $29^{\prime\prime}$.  The component 1306\_3 at $z = 0.6050$ has uncertainties of $\sim38\%$ on the mass, $\sim19\%$ on the virial radius, and a centroid error of $43^{\prime\prime}$.   Contrastingly, 0850\_2 has statistical uncertainties of $\sim9\%$ on the mass, $\sim4\%$ on the virial radius, and a centroid error of $12^{\prime\prime}$.  As a result of this, the field magnification is better constrained in 0850 than in 1306.

%



\subsection{Constructing Magnification Maps}
\label{sect:magnification_maps}

We build maps of field magnification from mass models based on spectroscopy alone --- a complementary approach to lensing based methods.  We use the multi-plane tool GRAVLENS \citep{kee01} and a modified version of methodology by \citet{won11}.  In this model, we assign NFW \citep{nav97} halos to structures with velocity dispersions $\sigma > 300$ km s$^{-1}$.  NFW halo concentration is extracted from a mass-concentration relation derived by simulations by \citet{zha09}.  The fraction of the virial mass apportioned to halo mass is fixed at $1 - 0.07\times C$ (where $C$ is the redshift success completeness), as is assumed for Abell 1689 by \citet{lim07}.  The remainder is assigned to cluster member galaxies.  We assign truncated singular isothermal spheres (SIS) to galaxies not associated with group or cluster halos, with a truncation radius of $r_{200}$.  We also use truncated SIS halos to model galaxies flagged as members of groups or clusters.  The group galaxies' truncation radii are scaled so that the density at the truncation radius is the same for all group galaxies \citep[see equation B4 in the paper by][]{won11}.  As a result the truncation radius for any group member depends on the total mass assigned to the galaxies for that parent halo.  Halo masses for individual galaxies (both group members and non-members) are derived from their absolute magnitudes using the Faber-Jackson relation between absolute magnitude and velocity dispersion \citep{fab76}.  All clusters and groups, as well as galaxies within 3 arcminutes of the field center, are treated explicitly; all other galaxies are treated with the shear approximation \citep{won11, mcc13}.    

In order to produce a mass model that accurately reflects our uncertainties in measurements as well as halo parameters like ellipticity that are difficult to constrain, we construct an ensemble of mass models with a Monte Carlo simulation and report descriptions of the aggregate rather than individual fiducial models.   For this ensemble, we vary halo concentration, halo ellipticity, halo centroid, total mass, and the properties of individual galaxies over 1000 trials using GRAVLENS.  We construct suites of mass models for three source plane redshifts:  $z=5.03$ (to better match the best-fit photometric redshift of the candidate multiply-imaged source described in Section \ref{sect:arc_analysis}), $z=2.5$ (to enable direct comparison with published critical curves for MACS 1206-0847 \citep{ebe01}), and $z=10$ (to estimate the lensing properties of the beams for the purpose of detecting very high-redshift galaxies).  We vary the halo concentration according to the observed scatter in the mass-concentration relation \citep[0.14 dex,][]{bul01}.  We vary the halo centroid according to a bootstrap resampling of the observed galaxy positions.  The simulation varies the 3D axis ratios and orientations of the halos and projects these to the observed line of sight to generate a realistic ellipticity distribution, as in the papers by \citet{sha06} and \citet{won12}.  Halo mass apportionment is varied between $1 - 0.01\times C$ and $1 - 0.1\times C$ \citep{gao11}.  The total halo mass, as constrained by the observed velocity dispersion, is also varied according to a bootstrap resampling of the galaxy positions and redshifts.  The properties of individual galaxies are varied according to scatter in the Faber-Jackson relation \citep{ber03} and the known photometric errors.  The result of the Monte Carlo is a distribution of 1000 magnification maps and traces of the tangential critical curve.

A useful measure of the utility of a beam for magnifying distant sources is the \'{e}tendue $\sigma_{\mu}$, or the areal coverage with magnification greater than a specified threshold.  This quantity is calculated in the source plane, as defined by \citet{won12}, and refers to the magnification of the brightest image of a source.  We also calculate this parameter in the lens plane to enable quick comparison with literature values.  In this paper, we calculate $\sigma_{\mu}$ for two $\mu$ thresholds of 3 and 10.  Depending on the observational goals of follow-up programs in these beams, one or both of these measures will be important for selecting powerful cosmic telescopes.  $\sigma_{\mu}$ values measured with a lower $\mu$ threshold of 3 are indicative of the beam's areal coverage of intermediate magnification, whereas the $\sigma_{\mu}$ values measured with a higher $\mu$ threshold of 10 are indicative of the beam's areal coverage of higher magnification, which may be helpful for pushing to the faint end of the luminosity function.

\begin{figure*}[]

  \begin{center}
 \plotone{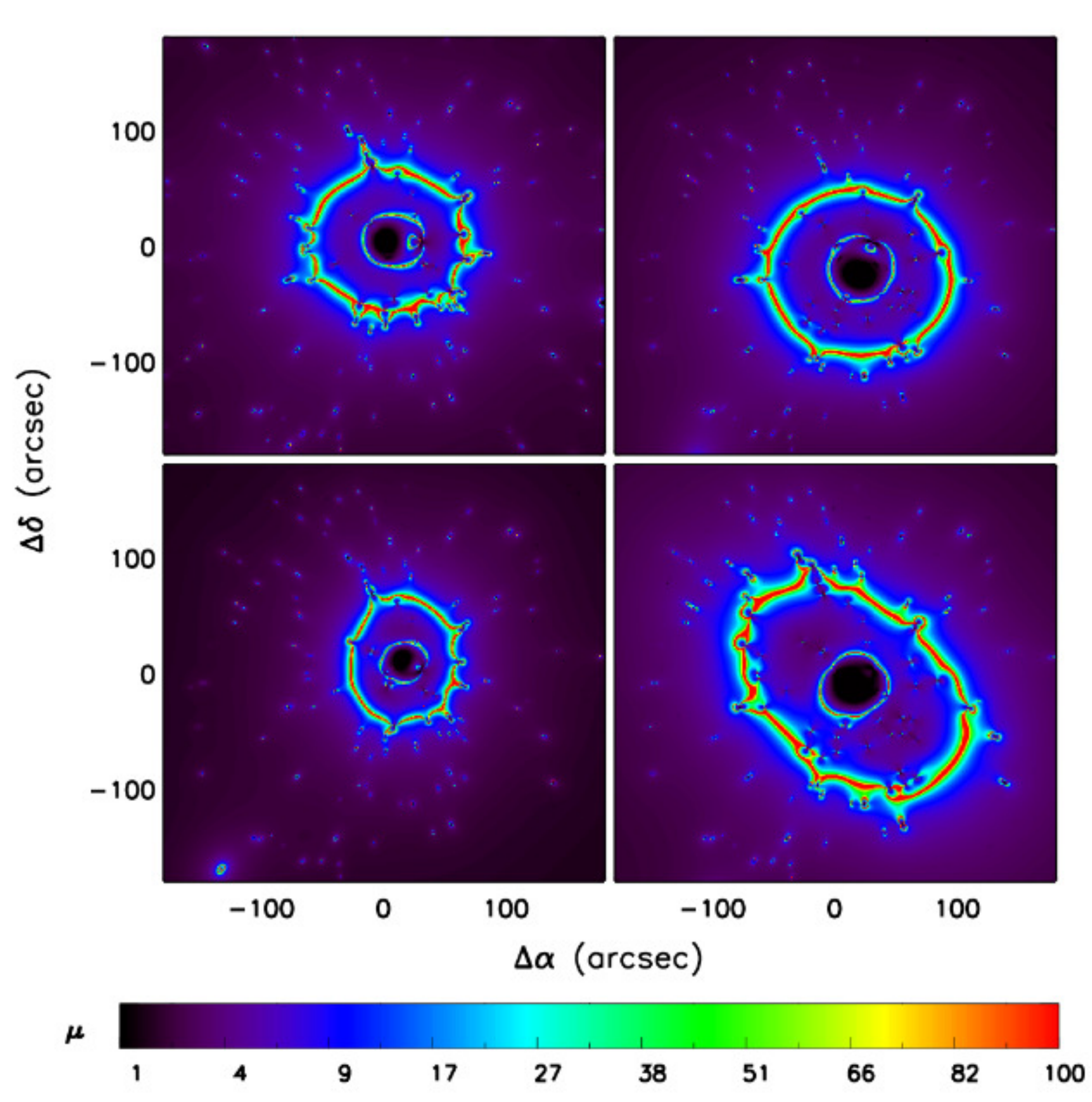}
  \caption{2D lens plane magnification maps in the field of beam 0850 for a $z=10$ source plane with a $360^{\prime\prime} \times 360^{\prime\prime} $ field of view.  The upper left panel shows a model with spherical halos that have halo mass, centroid, and radius exactly as measured from field spectroscopy (i.e., without Monte Carlo variation).  Individual galaxies not assigned to cluster-scale halos are included.  The upper right shows the map for the median Monte Carlo trial as ranked by $\sigma_{\mu}$ to a source plane of $z=10$ (calculated with a $\mu$ threshold of 10).  The lower left panel shows the map for a Monte Carlo trial ranked as the 15.8-th percentile (1$\sigma$) in $\sigma_{\mu > 10}$ to a source plane of $z=10$.  The lower right panel shows the map for a Monte Carlo trial ranked as the 84.2-th percentile (1$\sigma$) in $\sigma_{\mu}$ to a source plane of $z=10$.  The color bar scales logarithmically with magnification.  Although there is some difference between the 1$\sigma$ models (bottom panels), both possess substantial fields of high magnification.  Note that this information is drawn only from field redshifts and does not include strong-lensing information (We augment this model with strong lensing in Figure \ref{fig:0850_sigmamu_histogram}).  }
  \label{fig:0850_magmap_z10}
 \end{center}
\end{figure*}

\begin{figure*}[h]

  \begin{center}
 \plotone{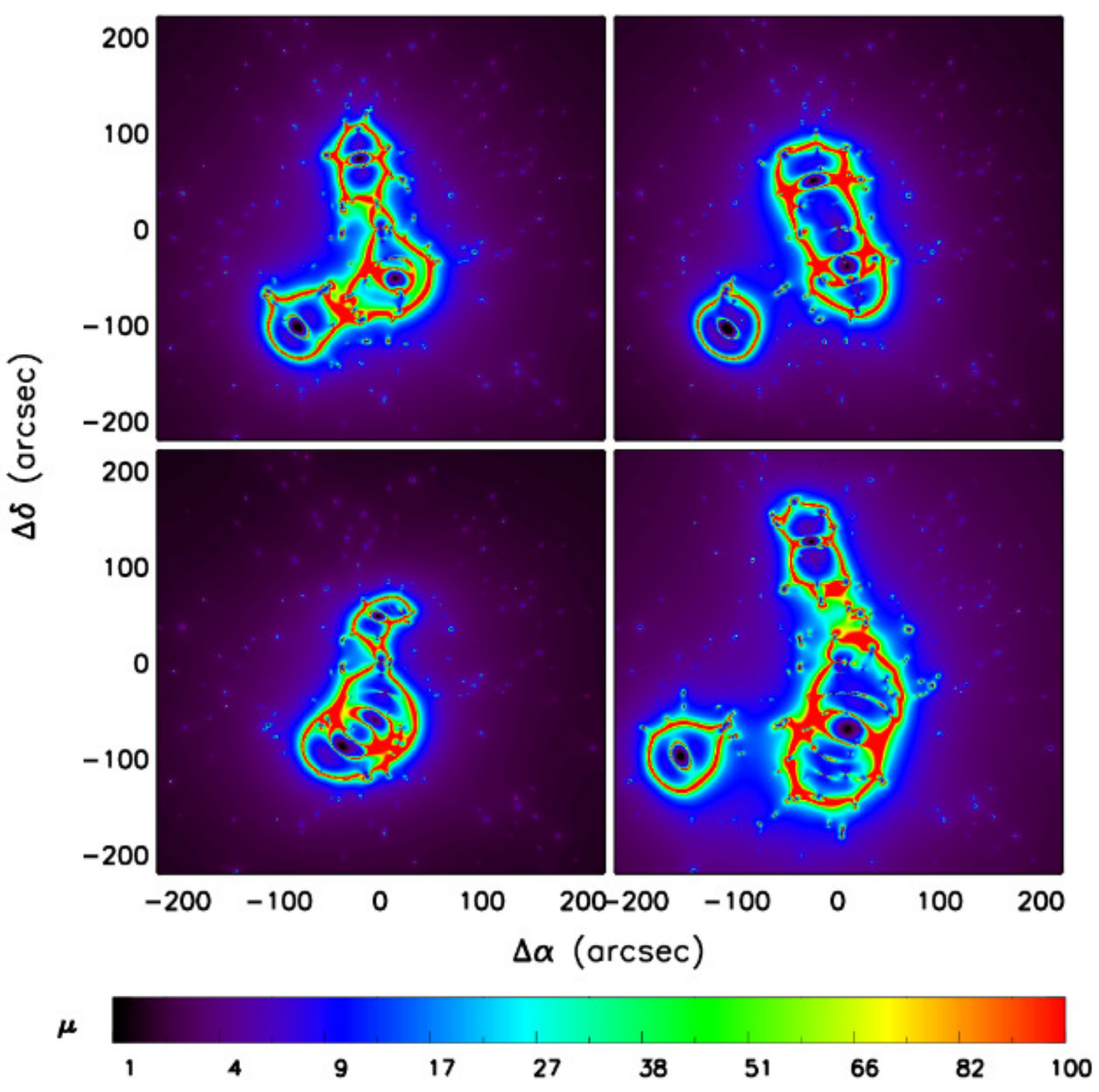}
   \caption{Same as Figure \ref{fig:0850_magmap_z10}, but for beam 1306.  A $440^{\prime\prime} \times 440^{\prime\prime} $ field of view is shown (slightly larger than in Figure \ref{fig:0850_magmap_z10} to include all structures in the frames).  Most Monte Carlo trials, and all of the panels shown here, display lensing interactions between the halos.  Note that there is some qualitative variation in the field magnification between the $1\sigma$ halos (lower left and lower right panels) as ranked by $\sigma_{\mu > 10}$ to a source plane of $z=10$.  This is a product of weaker constraints on the halo parameters of the two higher redshift components $1306\_2$ and $1306\_3$, especially the halo centroid, due to poorer galaxy sampling statistics.  However, the overall morphology of the lensed region is clear, and the total mass of the beam is statistically well-constrained ($3.4^{+0.4}_{-0.4}\times10^{15} M_{\odot}$).}
  \label{fig:1306_magmap_z10}
 \end{center}
\end{figure*}

For the new beams 0850 and 1306, we show sets of magnification maps for individual trials in the Monte Carlo ensembles for a source plane of $z=10$ in Figures \ref{fig:0850_magmap_z10} and \ref{fig:1306_magmap_z10}.  The upper left panel in both plots shows a model with spherical halos for which all other halo parameters (virial mass, centroid, and virial radius) are exactly as measured from field spectroscopy.  The upper right panel in both plots shows the median Monte Carlo trial for each beam as ranked by $\sigma_{\mu}$ (with a $\mu$ threshold of 10) to a source plane of $z=10.$  To give an indication of the variability of the models and the relative uncertainty in the halo parameters, we show the 15.8-th and 84.2-th percentile models as ranked by $\sigma_{\mu}$ in the bottom panels of Figures \ref{fig:0850_magmap_z10} and \ref{fig:1306_magmap_z10}.

In 0850 (Figure \ref{fig:0850_magmap_z10}), all halo parameters of the most massive component 0850\_2 are well constrained by field spectroscopy, so there is relatively little variation between the median, 15.8-th, and 84.2-th percentile ranked models.  Both of the $1\sigma$ models possess substantial fields of high magnification, suggesting a high probability that 0850 is a valuable beam for lensing high redshift sources.  We later show that this conclusion is consistent with the location of a candidate multiply-imaged galaxy at $z=5.03$ (Section \ref{sect:arc_analysis}).

\begin{figure*}[]
  \begin{center}
  	\epsscale{1.0}
 \plotone{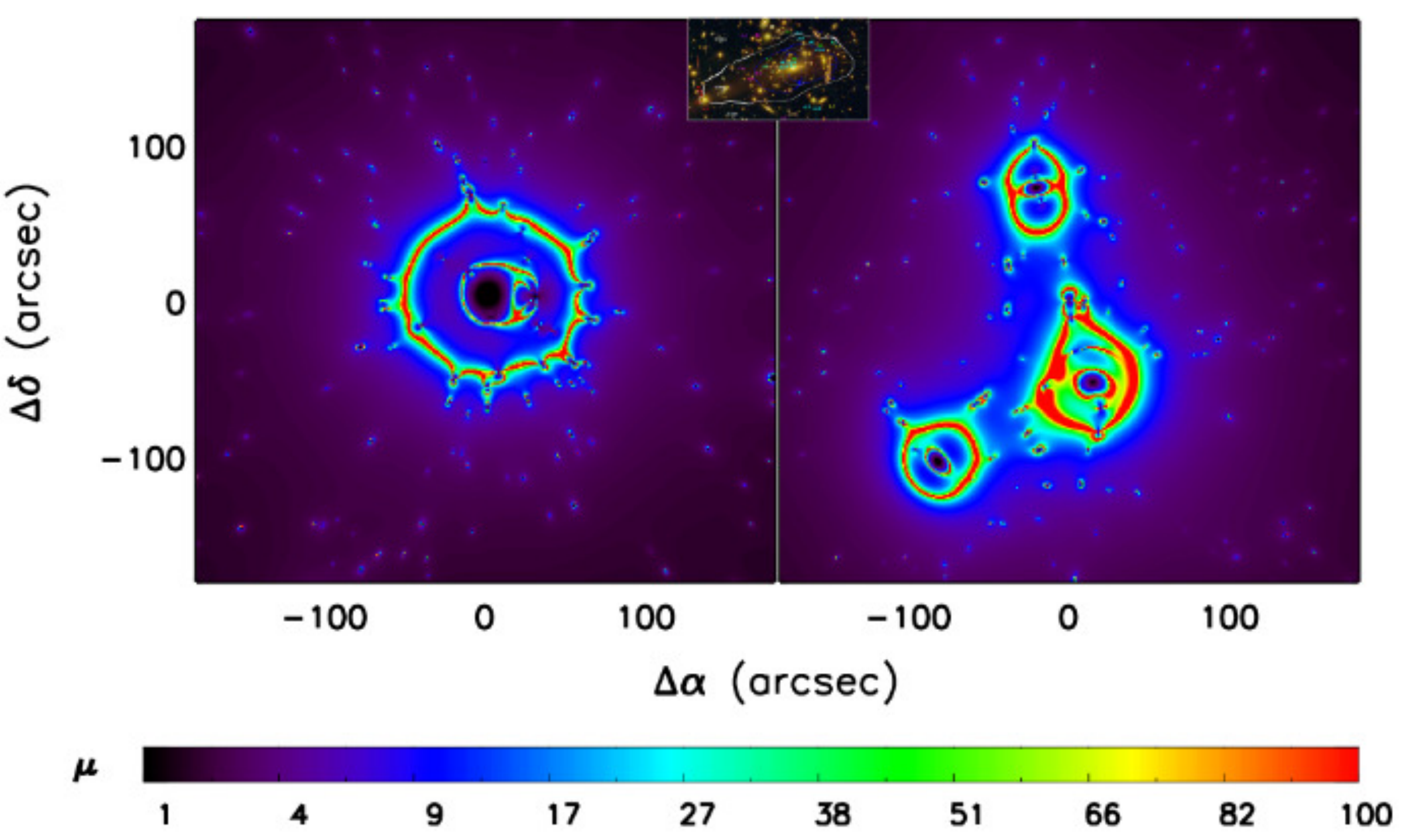}
  \caption{Comparison of magnification maps in the field of beams 0850 (left panel) and 1306 (right panel) for a $z=2.5$ source plane with those of MACS 1206-0847.  The field of view is $360^{\prime\prime} \times 360^{\prime\prime} $.  These models assume spherical halos and set virial masses, centroids, and virial radii to the exact values derived from galaxy spectroscopy and SDSS photometry.  An ACS/WFC3 image of MACS 1206-0847, one of the more massive ($\sim 1.5 \times 10^{15} M_\odot$) clusters in the CLASH survey, is included in the inset for comparison at the same spatial scale \citep{pos12}.  Its critical curve for $z_s = 2.5$ is shown in white \citep{zit12}.  These spherical models surpass MACS 1206-0847 in terms of the size of the magnified region available to lens distant galaxies.  The model for beam 1306 shows some lensing interactions between halos, which can boost the lensing potential beyond what is implied from the total mass alone \citep{won12}.    Figures \ref{fig:0850_magmap_z10} and \ref{fig:1306_magmap_z10} give a visual impression of the uncertainties in these magnification maps, although the source redshift for those figures is 10 rather than 2.5.}
  \label{fig:0850_1306_magmap}
 \end{center}
\end{figure*}

However, in 1306 (Figure \ref{fig:1306_magmap_z10}) there is some variation between the 15.8-th, and 84.2-th percentile ranked models as ranked by $\sigma_{\mu}$.  This is a direct result of poorer galaxy sampling in the two higher-redshift halos 1306\_2 and 1306\_3 at redshifts of $z = 0.3746$ and $z=0.6050$.  The variation in the exact morphology of the lensed region is largely due to uncertainty in the halo centroids of 1306\_2 and 1306\_3.  Although the magnification realizations are qualitatively very different in 1306, even the 15.8-th percentile ranked model possesses a considerable field of intermediate magnification and is qualitatively similar to the magnification produced by other well-known lensing clusters.  The 84.2-th percentile model has $6.7$ square arcminutes of area in the lens plane above a magnification threshold of 10 to a source plane of $z_s = 10$.  The total mass in the beam is statistically well-constrained with a measurement of $3.4^{+0.4}_{-0.4}\times10^{15} M_{\odot}$.

To enable visual comparison with published critical curves of known lensing clusters, we also generate magnification maps at a source redshift of $z_s = 2.5$.  For these models, we assume spherical halos and set other halo parameters (virial mass, centroid, and virial radius) to the exact values measured from field spectroscopy.  These magnification maps are shown in Figure \ref{fig:0850_1306_magmap} for beams 0850 and 1306.  Note that these are ``fiducial'' models that do not give a sense of the uncertainties in the magnification due to either measurement error or halo ellipiticity uncertainty.   For comparison, we include an ACS/WFC3 image of MACS 1206-0847, one of the more massive ($\sim 1.5 \times 10^{15} M_\odot$) clusters in the CLASH survey \citep[Cluster Lensing And Supernova survey with Hubble; ][]{pos12}.  In both beams, the spherical model qualitatively surpasses MACS 1206-0847 in terms of the size of the magnified region.

Table \ref{tab:tabulated_results_5} presents calculated values for $\sigma_{\mu}$ assuming a source redshift of $z_s = 10$ and two magnification thresholds, 3 and 10.  We present 68\% confidence intervals only, which are the 15.86\% and 84.14\% ranked values derived from the 1000 Monte Carlo trials.



\begin{table*}[ht]
\caption{Derived Lensing Properties of 0850 and 1306.}
\begin{center}
\leavevmode
\begin{tabular}[t]{cccc}

\tableline\tableline

\multicolumn{1}{c}{Parameter} &\multicolumn{1}{c}{Multiply-imaged arc }&\multicolumn{1}{c}{0850}    &\multicolumn{1}{c}{1306}  \\
\multicolumn{1}{c}{} &\multicolumn{1}{c}{as position constraint?} &\multicolumn{1}{c}{68\% confidence band}    &\multicolumn{1}{c}{68\% confidence band}  \\
\tableline	\\
\rule{0pt}{1ex} $\sigma_{\mu > 3}$ in source plane, $z_s = 10$ & no & [0.75, 2.4] &  [1.7, 2.6] \\
\vspace{0.06in}
$\sigma_{\mu > 3}$ in image plane, $z_s = 10$ & no & [4.2, 14] &  [9.4, 18] \\
$\sigma_{\mu > 10}$ in source plane, $z_s = 10$ & no & [0.06, 0.20] &  [0.11,0.27] \\
\vspace{0.06in}
$\sigma_{\mu > 10}$ in image plane, $z_s = 10$ & no & [1.2, 3.8] &  [2.3,6.7]\\
\tableline	\\
\rule{0pt}{1ex} $\sigma_{\mu > 3}$ in source plane, $z_s = 10$ & yes & [1.1, 2.6] & - \\
\vspace{0.06in}
$\sigma_{\mu > 3}$ in image plane, $z_s = 10$ &yes & [6.3, 15] & - \\
$\sigma_{\mu > 10}$ in source plane, $z_s = 10$ &yes & [0.10,0.22] & - \\
\vspace{0.06in}
$\sigma_{\mu > 10}$ in image plane, $z_s = 10$ &yes & [1.8, 4.2] &  - \\

\tableline\tableline
		  
\tableline
\end{tabular}

\label{tab:tabulated_results_5}
\end{center}
{\bf Notes.}  All $\sigma_{\mu}$ measurements in units of square arcminutes.
\end{table*}

\begin{figure*}[h]

  \begin{center}
 \plotone{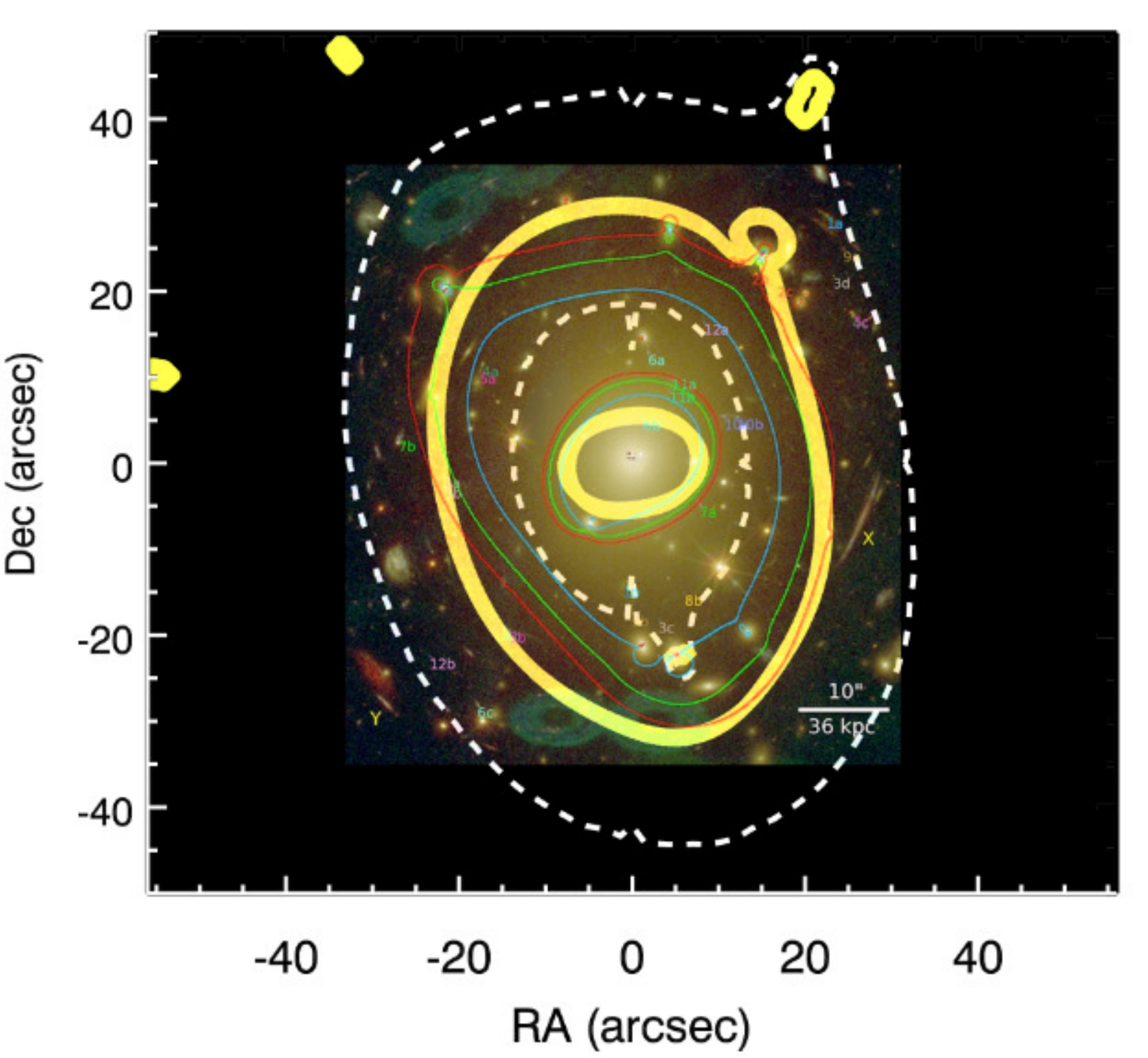}
  \caption{Comparison of Abell 2261 critical curves for a source redshift of $z_s = 2$ calculated from field redshifts.  The image is overlaid on Figure 1 of \citet{coe12}.  Our model is constructed using galaxy redshifts to estimate halo membership, halo mass, and virial radius.  Our model is augmented with some strong lensing-derived information; halo centroid, projected ellipticity, and position angle are taken from \citet{coe12}, in which they are derived from the cluster image using the ``mass follows light'' technique \citep{zit09b}.  Yellow bands mark the fiducial critical curves of our model and white dashed lines show the $1\sigma$ confidence intervals on the position of the tangential critical curve.   For comparison, we show the critical curves derived from strong lensing information alone \citep{coe12}.   The critical curves derived from spectroscopy and the cluster image agree with those derived only from strong lensing information to within the error bars.  }
  \label{fig:abell2261_comp}
 \end{center}
\end{figure*}
As a test of what our methodology can do, we compare our estimated critical curves for the well-known lens Abell 2261 to that derived from a standard lensing analysis \citep{coe12}.  For this comparison, we obtain field galaxy redshifts measured with MMT/Hectospec over a 30$^{\prime}$ diameter field of view from \citet{coe12}.  We assign halo membership to galaxies and calculate virial mass and virial radius for candidate halos as described in Section \ref{sect:virial_radii}.  Only one candidate peak at $z=0.2255$ survives our iterative procedure in this field, corresponding to Abell 2261's redshift of $z=0.225$ \citep{coe12}.  The calculated virial mass is $2.5^{+0.3}_{-0.3} \times 10^{15} M_{\odot}$ and the virial radius is $2.7^{+0.1}_{-0.1}$ Mpc.  We use a halo centroid, projected halo ellipticity (0.16), and projected position angle measured with the ``Zitrin Gaussian'' method \citep{zit09b,coe12}, effectively assuming that mass follows light.  

For this comparison, we also compute $1\sigma$ confidence contours of the tangential critical curve locations for Abell 2261.   These confidence contours are derived from a Monte Carlo ensemble of trials including realistic variation of all halo parameters, as described above.  For each of 360 rays emanating from the mean centroid of the most massive halo in 0850, evenly distributed in angle about 2$\pi$ radians, we record the radii of the intersection points of the predicted tangential critical curve with the ray for all Monte Carlo trials, including multiple intersections for individual trials if present.  The 68\% intervals in the distribution of radii for each ray define the positive and negative $1\sigma$ contours of the location of the critical curve for that angle.  

We show the fiducial critical curve for Abell 2261 and the $1\sigma$ confidence intervals on its location in Figure \ref{fig:abell2261_comp}.  The ``fiducial'' measurement uses a mass model with halo parameters exactly as measured, with no Monte Carlo variation.  The critical curves derived from spectroscopy and the cluster image compare favorably to those derived from strong lensing information only, and agree to within the substantial error bars.  The fiducial critical curve we construct is slightly larger than that from \citet{coe12}, likely due to the smaller halo mass of $2.2 \times 10^{15} M_{\odot}$ calculated in that study, as compared to the value of $2.5 \times 10^{15} M_{\odot}$ we calculate.  

The Abell 2261 test firstly indicates that if halo centroid and ellipticity are measured under comparable assumptions (i.e., mass follows light), a model constructed from redshifts predicts critical curve locations close to those predicted by a model derived from strong lensing.  Secondly, this test provides a check that NFW halo parameters are being treated correctly in our models of 0850 and 1306.  Note that we currently treat halo centroid, ellipticity, and position angle somewhat differently for the 0850 and 1306 models (halo centroid is measured from the member galaxies, and we marginalize about halo ellipticity and position angle in the Monte Carlo trials).  Constraining these halo parameters with the ``mass follows light'' assumption is a refinement upon our technique that could be performed with publicly available imaging and would further constrain the models shown in Figures \ref{fig:0850_magmap_z10} and \ref{fig:1306_magmap_z10}.

\subsection{Strongly-Lensed Arc Candidates in Subaru Imaging}
\label{sect:arc_analysis}

\begin{figure}[h]

  \begin{center}
 \plotone{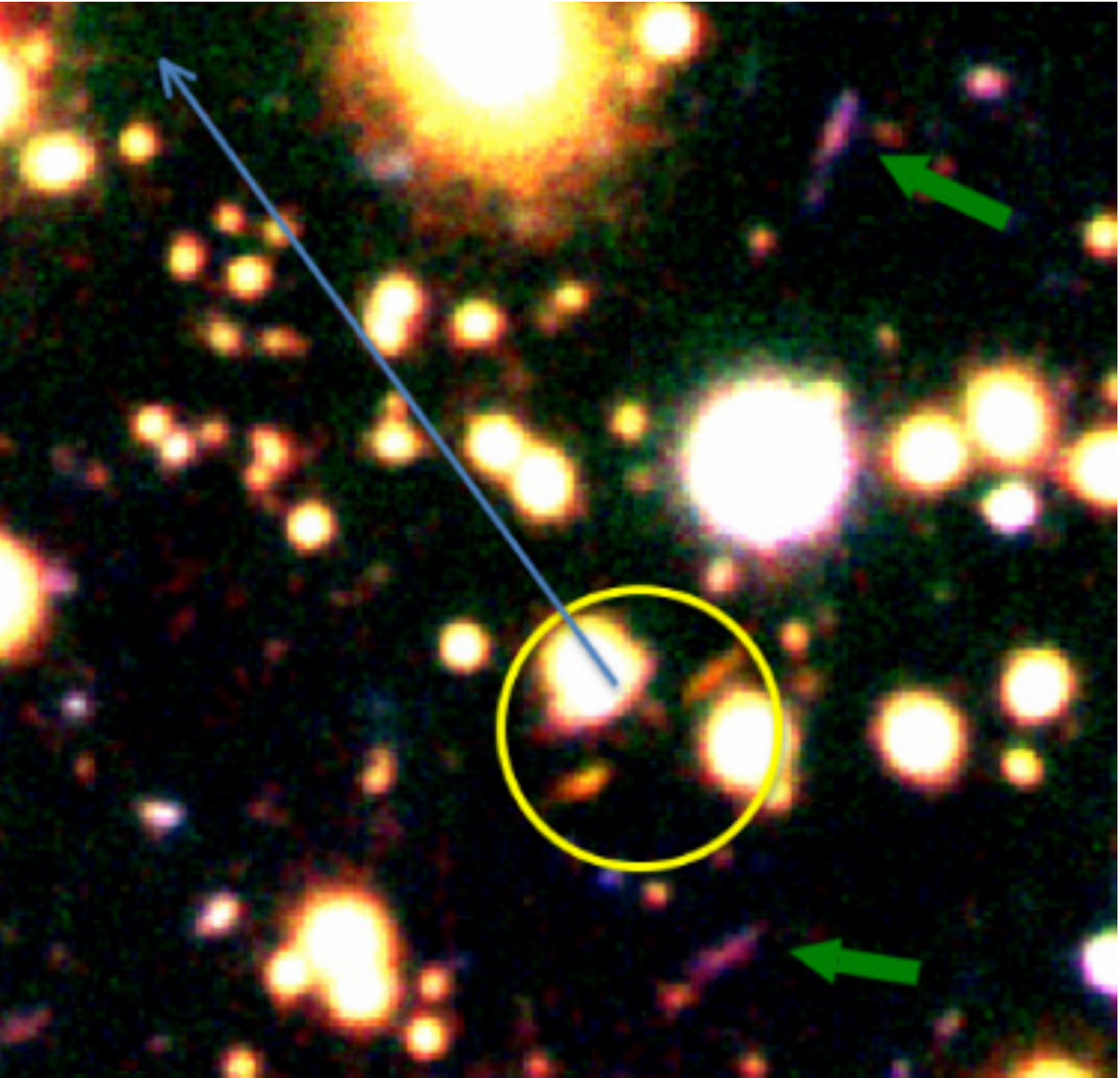}
   \caption{Subaru/Suprime-Cam color image ($1\farcm0 \times 1\farcm$0) containing the candidate
multiply-imaged galaxy in beam 0850, with components denoted by a yellow circle (arcs 2 and 3).  This image stacks and assigns $z^{\prime}$, $R_C$ + $I_C$ + $i^{\prime}$, and $B$ + $V$ to the RGB channels of the image.  Green arrows mark other candidate arcs in the field (arcs 1 and 4).
The blue vector points to the projected centroid of the most massive cluster in the beam as determined from spectroscopically confirmed members.  The morphologies, direction of elongation, and distance from the centroid of the dominant cluster ($45^{\prime\prime}$ - $55^{\prime\prime}$) suggest that they are highly magnified.}
  \label{fig:0850_arcs_color}
 \end{center}
\end{figure}

\begin{figure}[h]

  \begin{center}
 \plotone{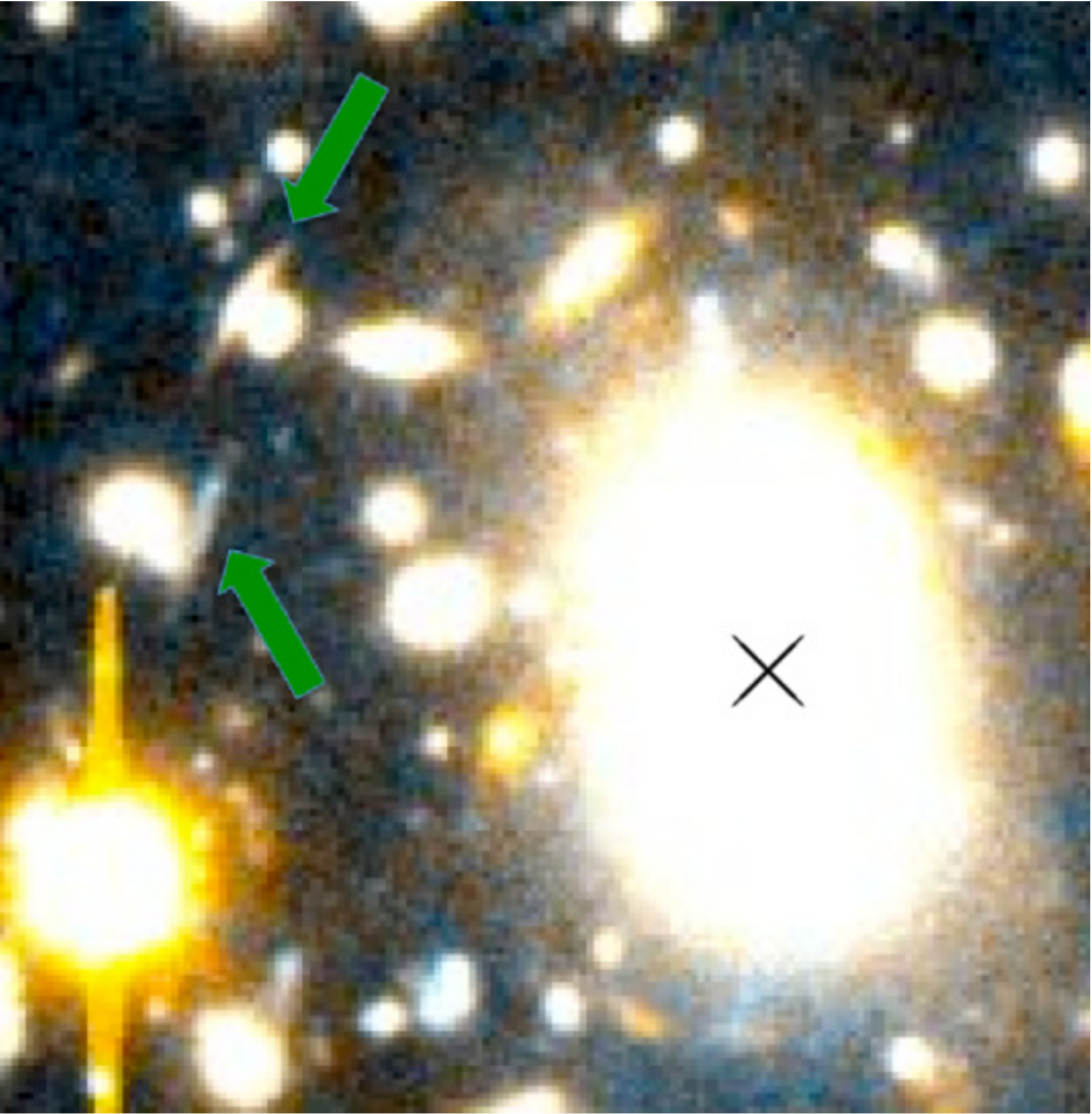}
  \caption{Subaru Suprime-Cam color image ($V$ and $i^{\prime}$, $42^{\prime\prime}\times42^{\prime\prime}$) showing new candidate strongly-lensed arcs in beam 1306, marked with green arrows.  The black X marks the position of the brightest cluster galaxy in the structure at $z = 0.2265$, 1306\_1.  \citet{san05} presents two arc candidates seen in WFPC2 imaging (not shown).  The two arcs identified here are detected in both the $V$ and $i^{\prime}$ bands.  Since neither are dropouts, conclusions about their source redshift are difficult.  However, their morphology and elongation perpendicular to the radial vector suggests that they are magnified.  }
  \label{fig:1306_arcs}
 \end{center}
\end{figure}

Multiply-imaged arcs are particularly valuable for constraining the mass distribution in multiple-cluster systems, as they ``pin down'' the location of the critical curves for a source plane at a given redshift.  We identify potential strongly-lensed arcs in both 0850 and 1306 in Figures \ref{fig:0850_arcs_color} and \ref{fig:1306_arcs} through visual inspection.  We have identified a candidate multiply-imaged galaxy in beam 0850 in deep multi-band Subaru imaging of this field (see Figure \ref{fig:0850_arcs_color}) through the morphology and color information of the two components.  The two adjacent components are both extended (with length-to-width ratio greater than 5), are individually tangential to the radial vector pointing to the center of the nearby massive cluster, are both V-dropouts with $V-I > 2 $, and have similar spectral energy distributions (SEDs).  The red $V-R$ color in both components is highly suggestive of a Lyman break at $z\sim5$,  but could potentially be the $4000\;$\AA\ Balmer break in a lower-redshift, dusty galaxy.  Their great distance from the center of the cluster (55$^{\prime\prime}$) suggests a large Einstein radius.  Thus we proceed to estimate their photometric redshift to test whether this beam is a massive cluster-scale lens.

 

\begin{figure}[h]
  \begin{center}
 \plotone{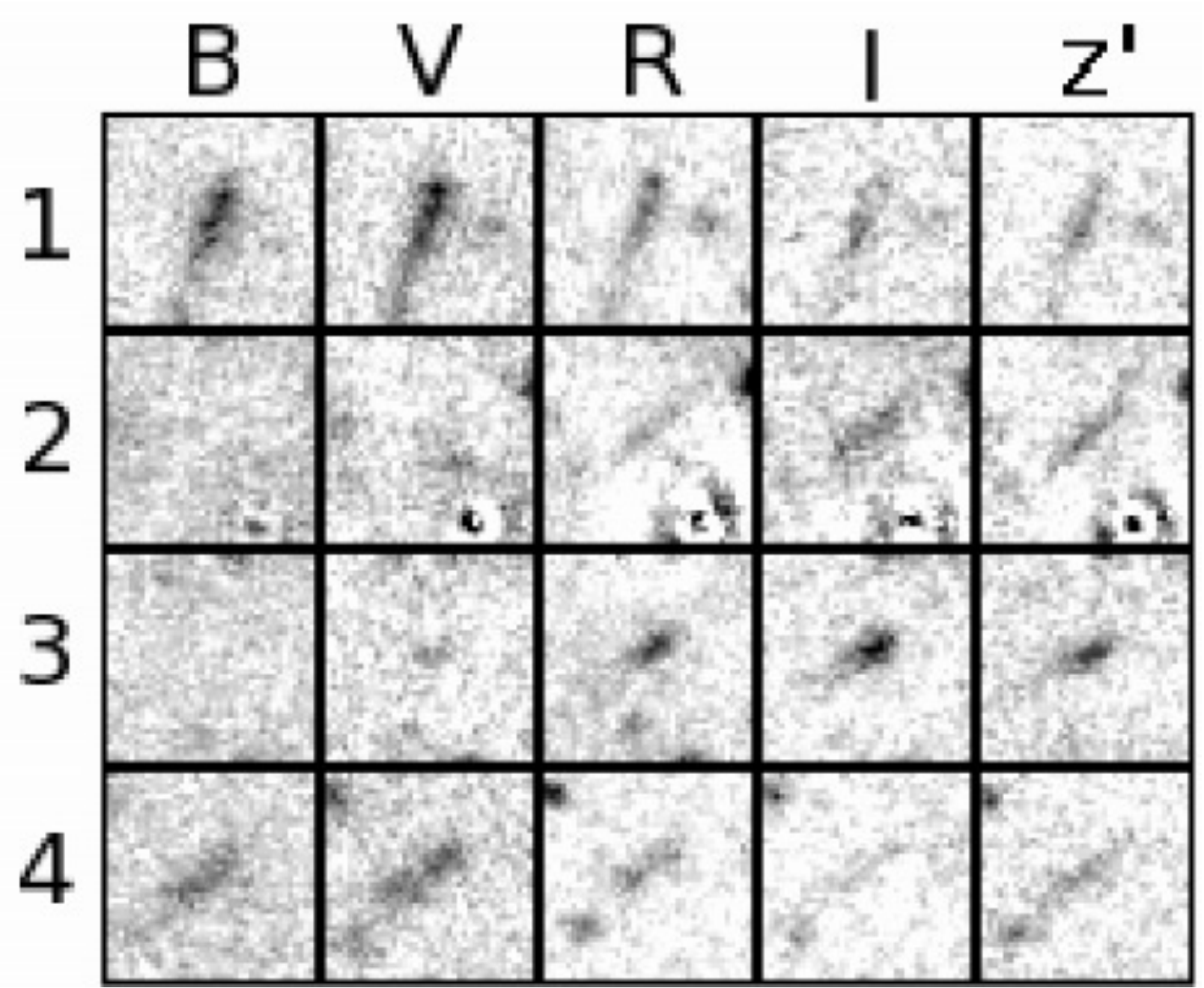}
  \caption{Subaru Suprime-Cam images of the four candidate strongly-lensed arcs in beam 0850, $8\farcs0 \times 8\farcs0.$  Each row denotes images of a different arc and columns show images in a different band, as labeled.  Nearby galaxies have been cleaned with GALFIT for arc 2 and 3.  Notice that arcs 2 and 3 are less visible in the B and V bands, suggesting that they are V-dropouts.  Arcs 1 and 4 are morphologically consistent with being lensed galaxies, but their detection in all bands suggests that they are at lower redshift ($z < 2$).}
  \label{fig:0850_arcs}
 \end{center}
\end{figure}

To assess whether these dropout arcs 2 and 3 (Figures \ref{fig:0850_arcs_color} and \ref{fig:0850_arcs}) are candidate multiply-imaged, we compare the SEDs and photometric redshifts for both.  Due to the potential contamination of arc photometry from lower-redshift galaxies, we use GALFIT \citep{pen02} to fit and remove adjacent galaxies before measuring photometry.  We fit elliptical 2-D sersic profiles to three galaxies in the $15^{\prime\prime}\times15^{\prime\prime}$ region surrounding arcs 2 and 3.  Residual images for each band for arcs 2 and 3 are shown in Figure \ref{fig:0850_arcs}.

We measure photometry on GALFIT residual images with elliptical apertures of axis ratio 2.8 and long axis $2\farcs8$, elongated in the direction of the extension of each arc.  We determine the sky background with annular elliptical apertures of axis ratio 2.8, inner radius $2\farcs8$, and outer radius $3\farcs5$.  Photometric zeropoints are measured by comparing to SDSS photometry of unsaturated stars as described in Section \ref{sect:subaru_data_reduction}.  We obtain error bars for each band by performing aperture photometry with the same elliptical apertures on $50$ non-overlapping blank regions in the $120\farcs \times 120\farcs$ region surrounding the center of the beam, computing the $68\%$ confidence intervals on the resulting flux distribution, and combining the result with the known zeropoint error for each band.  The positive $68\%$ confidence interval is quoted as the $1\sigma$ upper limit for non-detections.  These upper limits match those from the online Subaru Suprime-Cam Exposure Time Calculator to within $0.3$ magnitudes in all bands for photometric apertures of comparable area, assuming point sources and $0\farcs7$ seeing.

\begin{figure}[h]

  \begin{center}
 \plotone{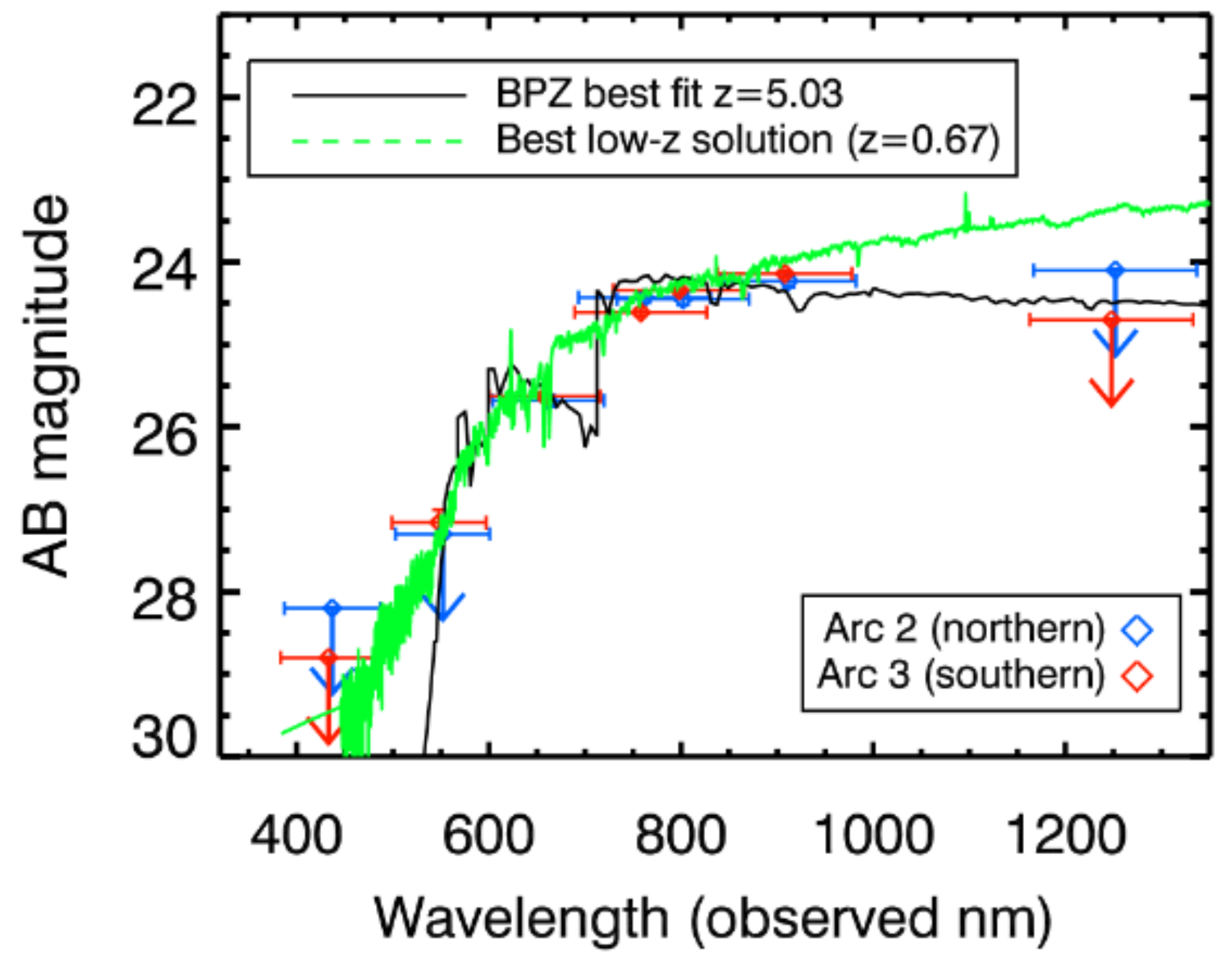}
  \plotone{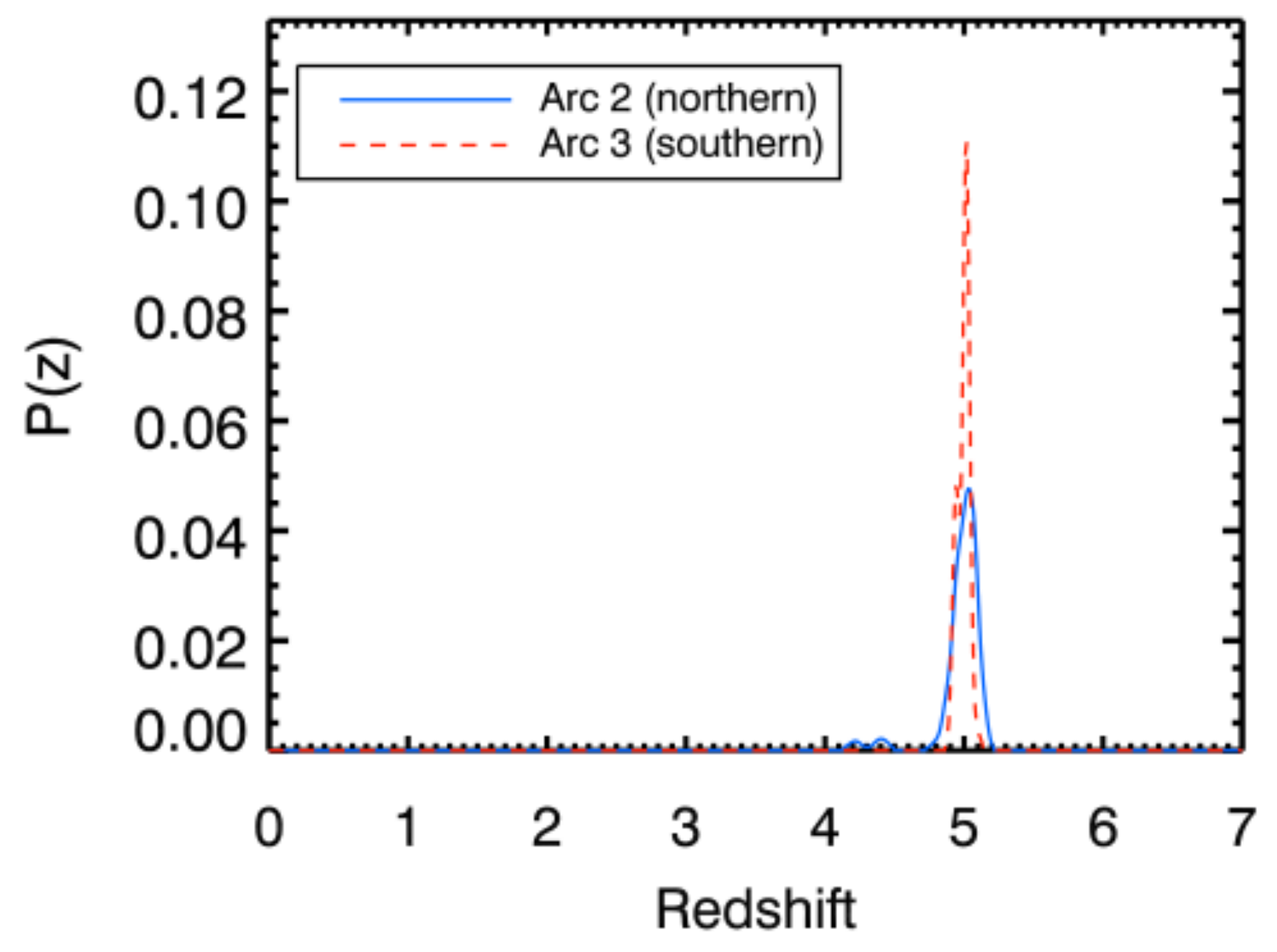}
  \caption{Top:  SEDs for the two candidate multiply-imaged V-dropout images in beam 0850 as a function of observed wavelength.  Blue diamonds mark the photometric points of arc 2 (northern) and red diamonds mark those of arc 3 (southern).   The photometry for arc 3 is shifted fainter by $0.6$ magnitudes for display purposes.  The B-band and J-band photometric measurements, as well as the V-band measurement for arc 2, are shown as $1\sigma$ upper limits.  The solid black line is the best-fit single stellar population (SSP) model at redshift $z=5.03,$ identifying the break at $\sim6000\;$\AA~as the Lyman Break.  The statistical error on the photometric redshift is $+0.21, -0.17.$   The best-fit low-redshift template solution for $z=0.7$ is shown as a dashed green line.  The majority of the discriminating power is provided by the $B-$ and $V$-band constraints.  Bottom:  BPZ photometric redshift probability distributions for both lensed arcs.  The probabilities of the high redshift solution being correct are $96.9\%$ for arc 2 and $99.3\%$ for arc 3.  This high probability, together with the low likelihood that two low-redshift interlopers would be found near each other with similar morphologies and SEDs, suggests that these arcs are multiple images of the same $z=5.03$ galaxy.}
  
   \label{fig:0850_SEDs}
 \end{center}
\end{figure}

Aperture photometry for 0850 arc candidates 2 and 3 is plotted in the top panel of Figure \ref{fig:0850_SEDs}.  Arcs 2 and 3 are not detected in the LBT/LUCI J-band image of 0850 at the $3\sigma$ level, so we plot one-sigma upper limits, measured using the same photometric apertures as for the optical photometry.  Excluding non-detections in the $B$ and $V$-bands, the photometry of the two arcs matches to within $1\sigma$ in all bands, suggesting that the two sources are multiple images of the same galaxy.

To be bona-fide multiply-imaged arcs, sources 2 and 3 would need to be located near the critical curve; but their significant distance of $55^{\prime\prime}$ away from the central BCG suggests that they must also be at high redshift ($z > 2$).  Here, we use the public photometric redshift code BPZ \citep{ben00} to estimate photometric redshifts for the two arcs independently.  This template-fitting code uses a Bayesian approach with priors of redshift probability as a function of galaxy type and magnitude.  For this analysis, we retain BPZ's default priors, which are generated for objects with spectroscopic redshifts in the Hubble Deep Field and judged superior to a flat redshift prior \citep{ben00}.  This is a conservative choice, as such priors reduce the probability of finding high-redshift solutions for bright sources, including sources made apparently bright by high magnification.  In addition to BPZ's eight template SEDs, we add the SED library distributed in the SDSS pipeline \emph{idlspec2d} used to fit SDSS spectroscopic redshifts \citep{aih11}.  The set of SEDs includes single stellar models gridded over metallicity and age (ranging from 5 Myr to 5 Gyr), galaxy eigenspectra with emission lines, and star forming galaxy templates.  We use the i$^{\prime}$ and $I_c$ filters as separate constraints.  

The bottom panel of Figure \ref{fig:0850_SEDs} plots the resulting redshift probability distributions for both arcs.  The most likely redshift is $z=5.03$ for arc 2 and $z=5.04$ for arc 3.  The $1\sigma$ confidence intervals are $4.84 < z < 5.13$ for arc 2 and $4.90 < z < 5.08$ for arc 3.  BPZ calculates the odds that a given redshift solution is correct by integrating the probability distribution $p(z|C,m)$ within a $0.27(1+z_B)$ interval centered on $z_B$ \citep{mob04}.  BPZ reports that the odds of the high redshift solution being correct are $96.1\%$ for arc 2 and $>99.9\%$ for arc 3.  The best fit template SED for arc 2 is plotted in the top panel of Figure \ref{fig:0850_SEDs}.  For comparison, the best fit low-redshift solution (calculated by running BPZ with a maximum redshift of 1) for arc 2 is plotted as a dashed green line ($z=0.695$).   The majority of the discriminating power comes from non-detections in B and J.  

\begin{figure}[h]
  \begin{center}
    \epsscale{1.17}
 \plotone{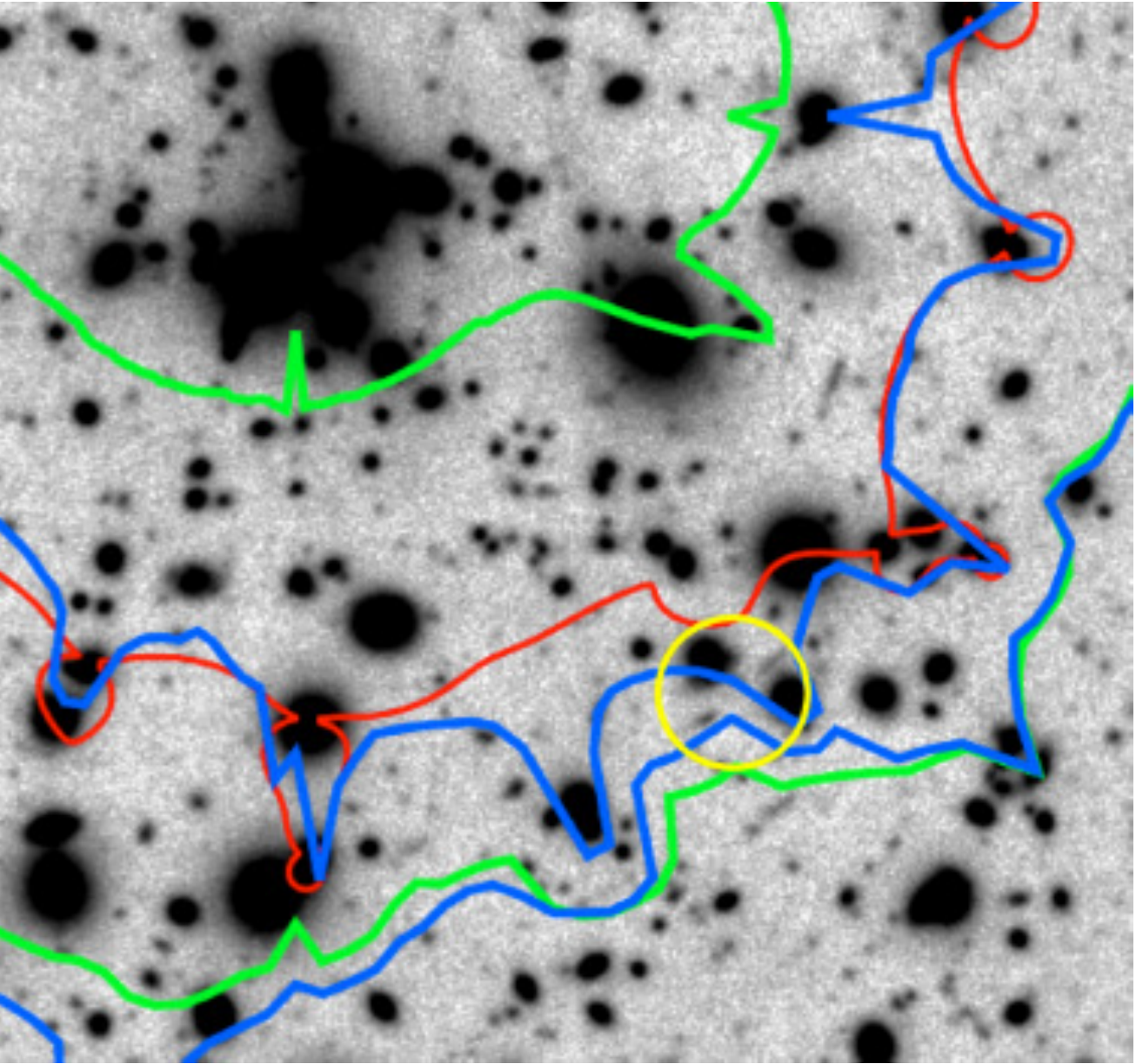}	
  \caption{Expanded view of the candidate multiply-imaged source in beam 0850 ($1\farcm6 \times 1\farcm$7).  Both components (yellow circle) are V-dropouts with a maximum-likelihood photometric redshift of $z = 5.03$.   Predicted critical curves for a ``fiducial'' model with spherical halos and halo mass, radius, and centroid exactly as measured from galaxy spectroscopy are shown in red ($z_s = 5.03$).  This model is the same as shown in the upper left panel of Figures \ref{fig:0850_magmap_z10} and \ref{fig:0850_1306_magmap}.   The green curves are positive and negative 68\% confidence intervals on the position of the critical curve, derived as described in Section \ref{sect:magnification_maps}.   The green confidence intervals do not necessarily enclose 68\% of the Monte Carlo trials; rather, along any given radial vector centered on the peak centroid of the most massive halo, 68\% of the critical curves intersect the vector between the green confidence intervals.  The location and orientation of the fiducial critical curves for this source plane, as determined by our mass model derived from galaxy spectroscopy and SDSS imaging alone, are consistent with the multiply-imaged source's location.  This agreement remains even when we use alternative measurements for the halo centroid (X-ray peak or BCG location).  Note that the addition of rudimentary strong-lensing information --- the constraint provided by the position of the multiply-imaged source --- further restricts the available range of the critical curves near its position (blue curves).}
   \label{fig:0850_arcs_CC}
 \end{center}
\end{figure}

The predicted critical curve for a source redshift of $z=5.03$ is shown in Figure \ref{fig:0850_arcs_CC}.  The critical curve is predicted from a mass model constrained by galaxy redshifts alone.  There is broad agreement between the location of the critical curve and the positions of the two arcs, suggesting that they are highly magnified.  We also make this comparison using two independent, alternative measurements of the centroid of the most massive halo:  the centroid of the X-ray peak and the Brightest Cluster Galaxy (BCG) centroid.  These centroids are within $15^{\prime\prime}$ of the cluster member centroid.  The agreement between the location of the predicted critical curve and the multiple images remains.  We also compute $1\sigma$ confidence contours of the tangential critical curve locations for beam 0850 as described in Section \ref{sect:magnification_maps} and present these as green lines in Figure \ref{fig:0850_arcs_CC}.   

The similar photometric redshifts of arcs 2 and 3 support the interpretation that they are multiple images of a $z\sim5$ galaxy.  However, the extreme similarity of the most likely values ($z=5.03$ and $5.04$) is a product of the particular shapes of the red SEDs and should not be taken as freestanding hard evidence of a relation between the two arcs, given the confidence intervals of $\sigma_z / z = 3.8\%$ for arc 2 and $\sigma_z / z = 3.2\%$ for arc 3.  However, a number of lines of evidence suggest that they are multiply-imaged, including the similarity of the photometric redshifts, SEDs, and morphologies, and their alignment with the predicted tangential critical curve in both distance from the cluster center and position angle of extension.

The locations, morphologies, and redshifts of strongly-lensed background galaxies are helpful constraints on the mass distribution in lensing galaxy clusters.  Multiply-imaged galaxies are particularly constraining because they fix the location of the tangential critical curve to lie roughly in between the multiple images.  Here, we incorporate the position and photometric redshift of the candidate multiply-imaged galaxy at $z\sim5.03$ into our estimates of the field magnification in 0850 without invoking sophisticated model fitting techniques.  Instead, given that our ensemble of 1000 Monte Carlo trials represents our best estimate of the real mass distributions, we simply select the trials that produce tangential critical curves that pass within $2^{\prime\prime}$ of the average position of the two components of the multiply-imaged galaxy (arcs 2 and 3 in Figure \ref{fig:0850_arcs}).  We treat this new ensemble of 130 trials as an estimate of the beam mass distribution as constrained by the position of the multiply-imaged galaxy.  Our match radius of $2^{\prime\prime}$ is chosen to produce a reasonably large set of matches with good number statistics and to reflect our uncertainty in the exact geometry of the tangential critical curve and the average centroid of the images (the separation between the two components is $8^{\prime\prime}$).

For the ensemble of 130 trials that produce critical curves passing within $2^{\prime\prime}$ of the average position of the components of the candidate multiply-imaged galaxy, we compute $1\sigma$ confidence contours of the tangential critical curve locations as described above.  These contours are plotted in Figure \ref{fig:0850_arcs_CC} as blue curves.  These may be compared with the confidence contours of the critical curves of the mass model derived from spectroscopy and SDSS imaging alone (green curves).  It is clear that the inclusion of strong-lensing information restricts the available range of the critical curves.  The range is most constrained in the immediate vicinity of the multiply-imaged galaxy (marked by a yellow circle).  

\begin{figure}[h]
  \begin{center}
  \epsscale{1.2}
 \plotone{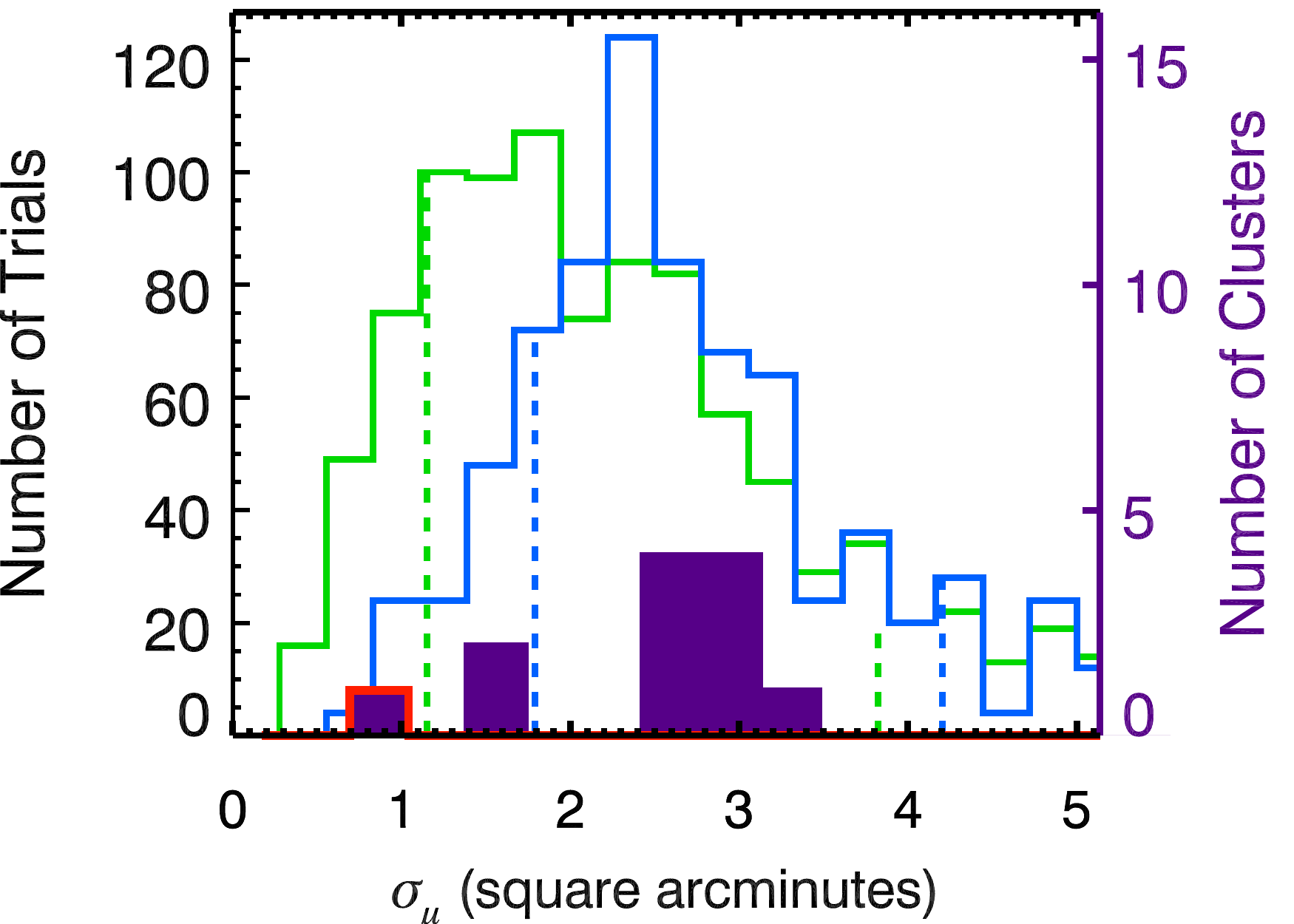}	
  \caption{Histograms of $\sigma_\mu$ values measured in the lens plane for a source redshift of $z=10$ and a $\mu$ threshold of 10 calculated for all 1000 Monte Carlo trials for beam 0850 (green curve).  The blue curve is the histogram for the 130 trials that produce critical curves passing within 2 arcseconds of the center of mass of the candidate multiply-imaged galaxy at $z=5.03.$  The dashed lines show the 68\% confidence intervals for both cases.  Note that the addition of the strong-lensing information has restricted the available range of  $\sigma_\mu$, excluding trials with low values that are unable to produce critical curves of sufficient radius to match the location of the multiply-imaged galaxy.  The addition of the strong-lensing information also increases the median value of the distribution.  For comparison, we include a histogram of $\sigma_\mu$ values measured in the lens plane for $\mu > 10$ and a source plane of $z_s = 8$ for 12 massive, X-ray luminous clusters at $z > 0.5$ in the MACS survey \citep[violet shaded region,][]{zit11}.  We also mark the $\sigma_\mu$ value measured in the lens plane for $\mu > 10$ and a source redshift of $z_s = 9$ for ACT-CL J0102-4915 \citep[``El Gordo'', red line,][]{zit13}.  These $\sigma_\mu$ values for source redshifts of $z_s = 8$ and $z_s = 9$ are very close (within 1\%) to the values that would be calculated for a source plane of $z_s = 10$ (A. Zitrin, priv. comm.), so they can be freely compared to the values we calculate for our beams with $z_s = 10$.  When including the candidate multiply-imaged galaxy at $z=5.03$ as a position anchor, the $\sigma_\mu$ confidence intervals for 0850 are comparable to or exceed the $\sigma_\mu$ values for El Gordo and the 12 other massive MACS clusters.  We note that the $\sigma_\mu$ range for 1306 is higher than 0850's range, although it is less tightly constrained.}
   \label{fig:0850_sigmamu_histogram}
 \end{center}
\end{figure}

Incorporation of strong lensing information into 0850's mass model produces tighter constraints on $\sigma_{\mu}$, or the areal coverage over a given magnification threshold.  We present the 68\% confidence intervals on $\sigma_{\mu}$ with a source redshift of $z_s  = 10$ in both the source plane and the lens plane for magnification thresholds of 3 and 10 in the bottom half of Table \ref{tab:tabulated_results_5}.   Note that for each perturbation of the $\sigma_{\mu}$ parameter, use of strong lensing information both narrows the available range of the parameter, and increases its median, 15.8\%-th, and 84.2\%-th percentile ranked values.  This information is shown graphically in Figure \ref{fig:0850_sigmamu_histogram}, which displays histograms of $\sigma_{\mu}$ for the Monte Carlo ensembles as measured in the lens plane for a source redshift of 10 and a magnification threshold of 10.  The green line in Figure \ref{fig:0850_sigmamu_histogram} is for the set of 1000 Monte Carlo trials using information derived from spectroscopy and SDSS imaging only; the blue line incorporates strong-lensing information from the candidate multiply-imaged galaxy at $z = 5.03.$  The 68\% confidence bands are shown as dashed lines for both cases.  Adding the strong lensing information excludes a number of trials with low values of $\sigma_{\mu}$ that are unable to produce critical curves at a radius sufficient to match the position of the images, and correspondingly, the median of the distribution increases.  

Figure \ref{fig:0850_sigmamu_histogram} also shows a histogram of $\sigma_{\mu}$ values for 12 massive, X-ray luminous clusters at $z > 0.5$ in the MACS sample, calculated in the lens plane for a source redshift of $z_s = 8$ and a $\mu$ threshold of 10 \citep{zit11}.  When including the candidate multiply-imaged galaxy at $z=5.03$ as a position anchor, the middle of the $\sigma_\mu$ range that we calculate for 0850 is comparable to these 12 MACS clusters.  That sample includes MACS 0717+3745, which has a considerable region of high magnification (3.5 square arcminutes in the lens plane) generated by a shallow, unrelaxed inner mass profile \citep{zit09a, zit11}.  Our confidence bands on $\sigma_{\mu}$ for 0850 and 1306 are comparable to MACS 0717+3745's \'{e}tendue:  [1.2, 3.8] and [2.3, 6.7] square arcminutes, respectively, calculated in the lens plane for a $z_s = 10$ source plane and using spectroscopy alone.  0850's $\sigma_{\mu}$ range is tightened to [1.8,4.2] square arcminutes after adding only one multiply-imaged source constraint.  Our $\sigma_\mu$ values are also comparable to or exceed that of the massive cluster ACT-CL J0102-4915 \citep[dubbed ``El Gordo'';][]{mar11, wil11, men12, has13}; \citet{zit13} measures a source plane area of $0.95$ square arcminutes for a $z_s = 9$ source plane for that cluster.  We plot this value as a red line in Figure \ref{fig:0850_sigmamu_histogram}.

\section{Conclusions}

\label{sect:conclusions}

In this study, we present 1151 galaxy redshifts in the fields of two lines of sight selected from the Sloan Digital Sky Survey \citep[SDSS DR9;][]{ahn12} for their high total luminosity densities of Luminous Red Galaxies \citep[LRGs; e.g.,][]{eis01}, 0850 and 1306.  We assemble group catalogs from the spectroscopy and identify five groups and cluster-scale halos in these two beams.  The four of five of these structures surpassing the group scale (with velocity dispersion greater than 500 km s$^{-1}$) are also traced by LRGs.  These two beams reflect the diversity seen in the \citet{won13} sample:  beam 0850 is dominated by a single massive halo and beam 1306 is composed of multiple cluster-scale halos.  Overall, in both 0850 and 1306, 16 of 18 LRGs (89\%) with spectroscopic redshifts are associated with halos that survive our iterative procedure and 17 of 18 LRGs (94\%) are associated with at least a visible peak seen in the galaxy redshift distribution.  This suggests that LRGs are a reliable tracer of groups and clusters.

We estimate the virial radii and masses for these halos following established techniques by \citet{zab90} and \citet{gir98}.  The total masses of these beams are substantial, $3.2^{+0.3}_{-0.3}\times10^{15} M_{\odot}$ for 0850 and $3.4^{+0.4}_{-0.4}\times10^{15} M_{\odot}$ for 1306.  For 1306\_1, the only halo with an X-ray-derived mass, the agreement with our spectroscopically-derived mass is excellent --- $M_{200} = 1.8 - 1.9 \times 10^{15} \Msun$ from X-ray \citep[after correcting from $M_{500}$ to $M_{200}$;][]{man10,rei11} versus $M_{200} = 2.0^{+0.19}_{-0.18} \times 10^{15} \Msun$ using our spectroscopy and SDSS imaging only.  We construct mass models for these beams using the spectroscopy only.   The resulting magnification maps suggest substantial fields of high magnification.  The morphology of the magnification distribution is well-constrained in 0850 and somewhat less well-constrained in 1306 due to poorer galaxy sampling; the addition of lensing data from the Hubble Space Telescope ({\it HST}) imaging and/or high quality ground-based imaging would greatly narrow the range of models in 1306, as we show in 0850.    

We also present a new serendipitous candidate multiply-imaged source at $z=5.03$ in 0850 seen in archival deep Subaru Suprime-Cam imaging.  Both components of the multiply-imaged source are V-dropouts, with a significant probability ($> 97\%$) of being at $z=5.03$.   The multiple-imaging interpretation of these sources is strengthened by the similarity of the photometric redshifts, SEDs, and morphologies, and their alignment tangential to a vector pointing toward the most massive cluster in the beam.  Our critical curve position in 0850 for $z_s = 5.03$ is consistent with the location of the multiply-imaged candidate, validating our preliminary mass model.  

We then augment the preliminary model with this strong-lensing constraint by selecting an ensemble of Monte Carlo trials that produce critical curves near the arcs of the candidate multiply-imaged source.  This addition both restricts the available range and increases the median value of the \'{e}tendue $\sigma_{\mu}$ for these clusters, a measure of the area above a specified magnification threshold.  

Other clusters with large \'{e}tendue have been discovered.  MACS 1206-0847 is a member of the MACS \citep[Massive Cluster Survey; ][]{ebe01} sample and is the most massive cluster in the CLASH \citep[Cluster Lensing And Supernova survey with Hubble; ][]{pos12} sample with an estimated virial mass of $1.5 \times 10^{15} M_{\odot}$ \citep{pos12}.  The predicted critical curves for beam 0850 at a source plane of $z=2.5$ \citep{zit12} are larger than those produced by MACS 1206-0847, suggesting a larger region of high magnification in 0850.  

The areal coverage of high magnification in these beams, as calculated in the lens plane for a source redshift of $z_s = 10$ and a $\mu$ threshold of 10, is comparable to published literature values for massive single-cluster lenses \citep{zit09a,zit11,zit13}.  The 68\% confidence bands in $\sigma_{\mu}$ for a $z_s = 10$ source plane that we calculate for 0850 and 1306 in the lens plane when using our galaxy spectroscopy and SDSS imaging --- [1.2, 3.8] and [2.3, 6.7]  square arcminutes, respectively --- are comparable to the published values for 12 massive, X-ray luminous members of the MACS sample.  When including the candidate multiply-imaged galaxy at $z=5.03$ as a position anchor, the confidence intervals on 0850's \'{e}tendue ([1.8, 4.2] square arcminutes) are comparable to the published \'{e}tendue of 3.5 square arcminutes for MACS 0717+3745.  Our $\sigma_\mu$ values are also comparable or larger than that of the massive cluster ACT-CL J0102-4915 \citep[dubbed ``El Gordo'';][]{mar11, wil11, men12, has13}; \citet{zit13} measures a source plane area of $0.95$ square arcminutes for a $z_s = 9$ source plane for that cluster.  

The favorable comparison between the calculated $\sigma_{\mu}$ values for our beams and established lensing clusters such as MACS 0717+3745 and ACT-CL J0102-4915 --- as well as the consistent evidence from both galaxy spectroscopy and strong lensing that the total masses are high --- confirm the substantial lensing power of these beams.





\section*{ACKNOWLEDGEMENTS}	

Thanks to Chang You and Decker French for assistance in gathering MMT Hectospec data and to Fuyan Bian for assistance in obtaining and reducing LBT/LUCI imaging.  Thanks to Adi Zitrin for providing $\sigma_\mu$ values for other clusters.  Thanks also to Dan Coe, Adi Zitrin, Dan Marrone, Brant Robertson, Sandy Faber, Pascal Oesch, Greg Walth, Michael Schneider, and Kristian Finlator for productive conversations.  Support for Program number HST-HF-51250.01-A was provided by NASA through a Hubble Fellowship grant from the Space Telescope Science Institute, which is operated by the Association of Universities for Research in Astronomy, Incorporated, under NASA contract NAS5-26555.   AIZ, CRK, and KCW acknowledge support from NASA through programs NNX10AD47G and NNX10AE88G and NSF support through AAG programs 1211874 and 121385.  KCW is supported by an EACOA Fellowship awarded by the East Asia Core Observatories Association, which consists of the Academia Sinica Institute of Astronomy and Astrophysics, the National Astronomical Observatory of Japan, the National Astronomical Observatory of China, and the Korea Astronomy and Space Science Institute.  This work is based in part on data collected at Subaru Telescope and obtained from the SMOKA, which is operated by the Astronomy Data Center, National Astronomical Observatory of Japan.  Observations reported here were obtained at the MMT Observatory, a joint facility of the University of Arizona and the Smithsonian Institution. This research has made use of the SIMBAD database, operated at CDS, Strasbourg, France.  This paper uses data products produced by the OIR Telescope Data Center, supported by the Smithsonian Astrophysical Observatory. This work is performed under the auspices of the U.S. Department of Energy by Lawrence Livermore National Laboratory under Contract DE-AC52-07NA27344 with document release number LLNL-JRNL-609412.  

{}

\end{document}